# Novel inferences of ionisation & recombination for particle/power balance during detached discharges using deuterium Balmer line spectroscopy

K. Verhaegh[1,2], B. Lipschultz[1], B.P. Duval[2], A. Fil[1,3], M. Wensing[2], C. Bowman[1,3], D.S. Gahle[4,3], the TCV team[5] and the EUROfusion MST1 team[6]

[1] York Plasma Institute, University of York, United Kingdom
[2] Ecole Polytechnique Fédérale de Lausanne (EPFL), Swiss Plasma Center (SPC), CH-1015 Lausanne, Switzerland
[3] Culham Centre for Fusion Energy, Culham Science Centre, OX14 3DB, UK
[4] Department of Physics SUPA, University of Strathclyde, Glasgow, G4 0NG, UK
[5] See the author list of S. Coda et al. 2019 Nucl. Fusion, accepted
[6] See the author list of B. Labit et al. 2019 Nucl. Fusion, accepted

## Abstract
The physics of divertor detachment is determined by divertor power, particle and momentum balance. This work provides a novel analysis technique of the Balmer line series to obtain a full particle/power balance measurement of the divertor. This supplies new information to understand what controls the divertor target ion flux during detachment.

Atomic deuterium excitation emission is separated from recombination quantitatively using Balmer series line ratios. This enables analysing those two components individually, providing ionisation/recombination source/sinks and hydrogenic power loss measurements. Probabilistic Monte Carlo techniques were employed to obtain full error propagation - eventually resulting in probability density functions for each output variable. Both local and overall particle and power balance in the divertor are then obtained. These techniques and their assumptions have been verified by comparing the analysed synthetic diagnostic 'measurements' obtained from SOLPS simulation results for the same discharge. Power/particle balance measurements have been obtained during attached and detached conditions on the TCV tokamak.

## 1. Introduction
Divertor detachment is predicted to be of paramount importance in handling the power exhaust of future fusion devices such as ITER [1]. Aside from target power deposition due to radiation and neutrals, the plasma heat flux ($q_t$ in W/m$^2$) is dependent on the divertor target ion flux density ($\Gamma_t$ in ions/m$^2$/s) and electron temperature ($T_t$ in eV):

$$q_t = \Gamma_t(\gamma T_t + \epsilon) \quad (1)$$

where $\gamma$ is the sheath transmission coefficient ($\gamma \sim 7$) and $\epsilon$ is the potential energy deposited on the target (13.6 eV for deuterium ion recombination into an atom), with the kinetic energy deposited being $\Gamma_t \gamma T_t$. Crucial to the reduction of the heat flux is detachment [2-11], which involves a simultaneous reduction of $\Gamma_t$ and $T_t$. This is in contrast to 'attached' divertor operation [2-11] where $\Gamma_t$ increases while $T_t$ drops, limiting the possible target heat flux decrease (eq. 1). The reduction in ion target flux in the transition to detachment is thus a key element of detachment and forms one of the most easily observed detachment indicators.

Particle, power and momentum balance are interconnected and determine the relation between the ion target current and the temperature. They are thus key to detachment and therefore, measuring sinks and sources of particles, power and momentum are key to detachment. Important is that the power sinks for impurity radiation and hydrogenic radiation are separated, as they play different roles in the detachment process ([11, 12] and as shown below). In this work, we have improved the analysis of the Balmer series to obtain measurements not obtained previously: a full power/particle balance measurement of the outer divertor with separated hydrogenic and impurity radiation estimates.

Particle balance dictates that the ion target current ($I_t$ – equation 2) is due to the sum of ion sources in the divertor (ionisation – $I_i$) minus the sum of ion sinks in the divertor (recombination – $I_r$) [9, 13, 14] plus the net influence of any flow of ions from outside the divertor into the divertor ($I_{SOL}$, which can be positive or negative). A key realisation is, however, that the ion target current is generally considered 'self-contained' in the divertor [5, 14]: $I_{SOL}$ is expected to be negligible compared to recycling flux and thus $I_t$. This emphasizes the need for simultaneous ionisation/recombination measurements in the divertor. In this formulation of equation 2, the ion target current $I_t$ (ion/s) represents the target ion flux density $\Gamma_t$, integrated along the target's surface: $I_t = \int \Gamma_t$.

$$I_t = I_i - I_r + I_{SOL} \qquad (2)$$

The existence of volumetric recombination has been confirmed experimentally [9, 15-19] previously and is routinely monitored through qualitative measurements (such as line ratios) on tokamaks. It is in high density regimes sometimes found, through quantitative analysis, to be significant in the reduction of the ion target flux [9, 16-20]. Partially due to that, volumetric recombination is often expected to play a central role in target ion flux reduction [21-26]. However, in previous work on TCV [6], the volumetric recombination rate was shown to be only a small fraction of the reduction of ion flux, which is in agreement with recent TCV simulations [27] as well as $N_2$-seeded discharges in C-Mod [6]; further emphasizing the need for ionisation measurements. u

Ionisation is the primary determinant of $I_t$ (equation 2) during attached operation and at the detachment onset. However, each ionisation event costs potential energy (13.6 eV – not including molecular dissociation) as well as radiated energy due to excitation preceding ionisation. The power flow into the recycling energy ($P_{recl}$) as well as this energy cost of ionisation ($E_{ion}$) [5, 9, 13, 14, 24] determines the maximum ion source (and thus $I_t$ possible), as shown in equation 3 [5, 9] where recombination is neglected for simplicity. Estimating both the ionisation source and hydrogenic radiative losses enable estimating $E_{ion}$ and thus provide a key parameter for studying detachment.

$$I_t = \frac{P_{recl}}{E_{ion} + \gamma T_t} \qquad (3)$$

Power and particle balance are thus intertwined and the available $P_{recl}$ must be compatible with the measured amount of ionisation. Such behaviour has been identified qualitatively through experiments [7] and in SOLPS simulations [13, 28, 29] / analytic model predictions [5, 9, 13, 14, 24]. Although experimental indications for power limitation are available (either from inferred ion sources [9], or from qualitative spectroscopic 'indicators' based on $D_\alpha$ [30]), one weakness of previous results is that quantitative information on both divertor power/particle sinks/sources was not available. However, this study and other recent parallel studies [31, 32] aim to provide quantitative information on ionisation during divertor detachment [31, 33]. These recent parallel studies [31, 32] have similar goals but differ in the solution method and require more diagnostic measurements, such as specialised VUV spectroscopy and recombination edge (365 nm) measurements.

In this work, we have improved the analysis of the Balmer series to obtain measurements simultaneous estimates on the electron temperature, ionisation source, recombination sink and hydrogenic power losses. That analysis is generally applicable (section 6) and has been applied to spectroscopic data from the TCV tokamak. There are various challenges to the interpretation/analysis of the Balmer line series through passive spectroscopy, which had to be alleviated to achieve this:

1) Balmer lines, during detached conditions, can arise due to a mixture of excitation and recombination emission. To analyse the Balmer line series quantitatively, it has been previously assumed the emission is either fully excitation or recombination dominated [9, 16-18, 20, 34, 35]. We show that this is not generally a good assumption for lower-n Balmer lines (e.g. n= 3 to 7->2) during detached scenarios – as both excitation and recombination emission may be important, although recombination dominant emission is often a good assumption for high-n Balmer lines ($n \geq 9$) [8, 9, 16, 17, 19, 20, 35].
2) The emission is not well localised: an emission profile exists along the line of sight. Different plasma species can emit at different locations along that line of sight and their recombination/excitation emission can occur at different places. Sometimes, however, the same temperature is attributed to these regions, for instance by assuming– a plasma slab model with a single temperature for the analysis of passive spectroscopic signals [6, 31, 34, 36, 37]. Implicitly, it is assumed then that the temperature at the excitation/recombination emission location is the same [34, 36, 37], which may not be necessarily true.
3) In literature, it is sometimes assumed that Balmer line ratios indicative of recombinative emission imply that the plasma is 'recombination dominant' [15, 37, 38]. The number of expected photons per recombination/ionisation is, however, much different for recombination and ionisation and can be strongly dependent on the electron density. As we will show, this means that the recombination rate can be significantly lower than the ionisation rate, despite Balmer line emission being dominated by recombination: recombinative emission is thus not necessarily indicative of a larger recombination than ionisation *rate*. To determine that, a quantitative analysis is required.
4) Atomic data can be highly non-linear; which complicates both inferring results from spectroscopic measurements accurately as well as a full uncertainty quantification of that process.

The improvements in the Balmer line series analysis/interpretation in this work has alleviated many/all the above challenges. First, the technique enables quantitatively separating both the atomic excitation & recombination contributions to Balmer line emission by using the ratio between two Balmer lines. This resolves the first point. This technique is insensitive to chordal integration effects as well as uncertainties in the neutral fraction. After the separation, each of the two emission components is analysed individually to provide quantitative values for ionisation/recombination along each viewing chord resolving the third point. As the excitation/recombination emission contributions are analysed separately, one can – to some degree – account for the fact that both emission regions can be at different locations of the line of sight. Separate temperatures for the two regions are determined. The second point is partially resolved as the full analysis is less sensitive to chordal integration effects. This analysis technique has been verified using a synthetic diagnostic approach on SOLPS data, confirming that the analysis is insensitive to chordal integration effects. All analysis is performed using a Monte-Carlo probabilistic approach, which enables a full uncertainty quantification despite the non-linearity in the atomic data, resolving the fourth point.

First, we will provide an overview of the analysis strategy in section 3 and an introduction to the diagnostic used in section 2. Each individual step of the analysis is sequentially highlighted in sections 4.1 (Stark broadening), 4.2 (separation of excitation/recombination emission), 4.3 (inferring recombination/ionisation rates & charge exchange to ionisation ratios), 4.4 (inferring power losses associated with recombination/ionisation). Section 5 highlights the probabilistic analysis technique used, while section 6 provides an in-depth validation of the analysis technique using a synthetic diagnostic approach on SOLPS simulated discharges [27]. First results of the analysis technique, indicating power limitation of the ionisation rate on TCV is shown in section 7. Further implications of the analysis and applicability to future devices is discussed in section 8, while a conclusion is provided in section 9. The main analysis code for this is available at [39].

## 2. Diagnostic overview and experimental setup

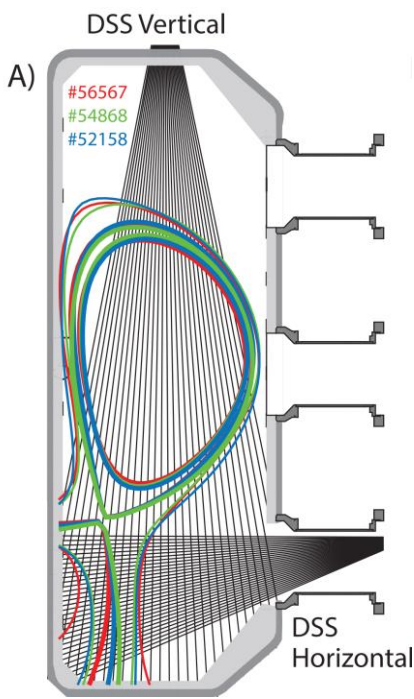

Figure 1. Lines of sight of the horizontal and vertical DSS systems. Divertor geometries for #56567 (red), #54868 (green), #52158 (blue) are shown. For the analysis in this work, only the horizontal DSS is used.

Although the shown analysis technique is more generally applicable, we will first introduce the diagnostic were this analysis has been applied to. A 'synthetic' version of this diagnostic has been developed (section 5) for SOLPS simulations to validate several aspects of the analysis (section 5).

The analysis technique has been applied to data from the newly developed TCV divertor spectroscopy system (DSS) [6]. The DSS consists of vertical and horizontal viewing systems, each employing 32 lines of sight (Figure 1). *Our analysis is based on the data from the horizontal viewing system*. Important for the analysis is that a full coverage for the divertor is obtained, which is true here for the outer divertor (spatial resolution of ~ 13 mm). The illustrated analysis may require calibrated (instrumental function calibration and absolute calibration) settings from several wavelength regions to obtain sufficient coverage/spectral resolution. That was available here as the spectrometer (Princeton Instruments Isoplane SCT 320) contains a triple grating turret which can be turned to change the grating used and to change the wavelength region covered (e.g. to enable measuring different Balmer lines). Further details on the DSS can be found in [10].

## 3. Balmer line analysis techniques

The aim of the analysis technique developed below is to provide a method for obtaining a full picture of the power/particle balance in the divertor using spectroscopy. First, we will start our description of the spectroscopic analysis with a brief review of our techniques and nomenclature for splitting excitation/recombinative emission / inferring recombination rates [6] as well as the organization of the analysis flow that is undertaken.

The brightness of a hydrogen Balmer line ($B_{n \to 2}$ in ph/m$^2$ s) with upper quantum number $n$ can be described (See Eq. 5) along a path length $\Delta L$ as function of electron density ($n_e$), neutral density ($n_o$) and temperature ($T_e$) using the Photon Emissivity Coefficients ($PEC_{n \to 2}^{rec}$) for recombination and excitation ($PEC_{n \to 2}^{exc}$), obtained from the Open-ADAS database [40-42] (in this work the following Open-ADAS data files were used: pec12_h.dat, scd12_h.dat, acd12_h.dat, plt12_h.dat, prb12_h.dat, ccd12_h.dat).

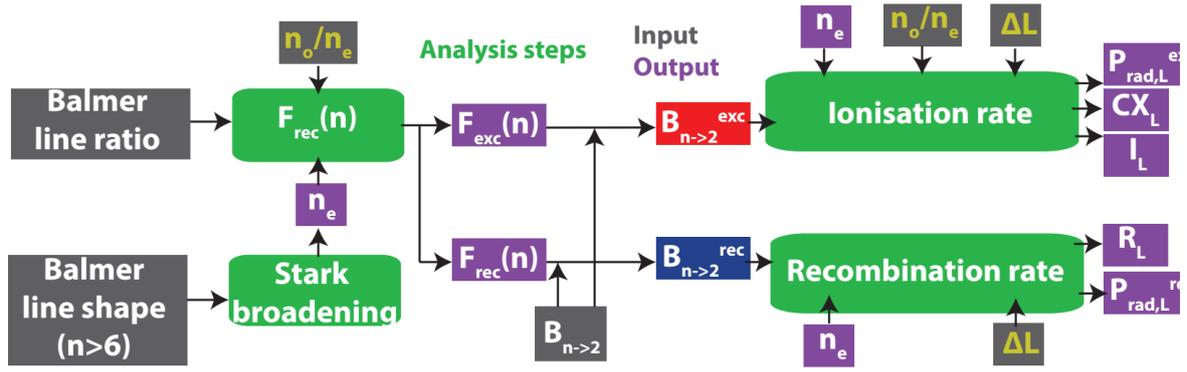

*Figure 2: Schematic overview of the analysis steps (green) in the Balmer line analysis chain together with the required measured inputs (grey – white text – including the Balmer line brightness $B_{n\to 2}$), assumed inputs (grey, yellow text – including the neutral fraction $n_o/n_e$, the path length $\Delta L$) and inferred outputs (purple – including the Stark density $n_e$, inferred recombination/excitation Balmer line emission fraction $F_{recl}(n)$, $F_{exc}(n)$; line integrated hydrogenic excitation/recombination radiated power loss $P_{rad,L}^{exc}$; $P_{rad,L}^{rec}$; line integrated ionisation/recombination rate $I_L$, $R_L$ and line averaged charge exchange to ionisation ratios $CX_L/I_L$).*

$$B_{n\to 2} = \underbrace{\Delta L\, n_e^2\, PEC_{n\to 2}^{rec}(n_e, T_e)}_{B_{n\to 2}^{rec}} + \underbrace{\Delta L\, n_e n_o PEC_{n\to 2}^{exc}(n_e, T_e)}_{B_{n\to 2}^{exc}} \tag{5}$$

Here it is assumed that: a) the Balmer line emission does not have molecular components (see section 3.5); b) the Balmer line emission originates from a plasma slab with spatially constant parameters (0D model) with a chord intersection length of $\Delta L$; c) the hydrogen ion density equals the electron density (e.g. $Z_{eff} = 1$) – which introduces insignificant errors on the analysis shown below [6].

Figure 2 illustrates the various steps in the analysis process, eventually resulting in estimates of both local plasma characteristics (weighted over the Balmer line emission profile along each viewing chord) and line integrated plasma parameters. The analysis starts with the Balmer line ratio and analysis of the Balmer line shape (Stark-broadened) to extract the density (section 3.1). These allow the determination of the fraction of the Balmer line brightness due to recombination and excitation (equations 6a and b) – section 3.2.

$$F_{rec}(n) = \frac{B_{n\to 2}^{rec}}{B_{n\to 2}} \tag{6a}$$

$$F_{exc}(n) = \frac{B_{n\to 2}^{exc}}{B_{n\to 2}} = 1 - F_{rec}(n) \tag{6b}$$

$F_{rec}(n)$ and $F_{exc}(n)$ are then combined with the absolute Balmer line intensity $B_{n\to 2}$ to obtain the absolute Balmer line emission due to recombination and excitation ($B_{n\to 2}^{rec}$, $B_{n\to 2}^{exc}$) (section 3.2). These are then analysed individually and independently (section 3.3) to obtain various local, line-integrated and toroidally integrated output parameters (section 3.3 & 3.7), including estimates on the recombination sink/ionisation source as well as the radiative power loss due to excitation and recombination. Those output parameters are summarised in section 3.7.

As shown in the flowchart – figure 2, several input parameters (e.g. $n_o/n_e$, $\Delta L$) are required and assumptions must be made to characterize them, described in section 3.6. The assumed uncertainty can be larger than 100% for some of those parameters. The effect is that a Taylor-expansion based error analysis is insufficient to accurately estimate uncertainties of the inferred output parameters. We thus developed and used a Monte-Carlo based probabilistic analysis to estimate output quantities and their uncertainties (section 4).

## 3.1 Stark broadening analysis for TCV discharges

The analysis chain starts with obtaining an estimate of the characteristic density of a chordal integral using Stark broadening, which represents an emission-weighted density along the line of sight. This has already been developed in [6] and has been improved further in here and [10] by performing a full Monte Carlo uncertainty analysis.

Considering characteristic TCV densities ($10^{19} - 10^{20}$ m$^{-3}$), the expected Stark broadening widths are relatively small and thus a high resolution setting (0.06 nm spectral resolution with 19 nm spectral range using a 1800 l/mm grating at a central wavelength of 404 nm) has been used to cover the line shape of the n=7 Balmer line (highest-n Balmer line which can be observed when only excitation emission is present on TCV) to determine the local density using simplified Stark models presented in [6, 43, 44]. Using the approach in [43], the Stark line shape is modelled as a 'modified Lorentzian' where the line shape is proportional to $\frac{1}{\left(\frac{\lambda-\lambda_0}{w}\right)^{5/2}+1}$. Here, $\lambda$ is the wavelength, $\lambda_0$ is the central wavelength of the spectral line and $w$ is a broadening parameter, given by tables in [43], dependent on electron density and electron temperature (which has a negligible contribution to $w$ [6] and is assumed to be 5 eV). Those models have been verified against a more complete Stark model utilizing a computer simulation technique to determine the Stark line shape [45] as shown in [46] using TCV DSS spectra. More information can be found in chapter 6 of [10] where this fitting technique has been compared in detail to fitting techniques based on a more complete Stark model [45, 46]; including discussions on the possible influence of Zeeman splitting as well as the possible influence of the electron temperature on the Stark profile.

The Stark-broadened width (the full-width-half-maximum) is small in the relatively low-density TCV plasmas compared to both the instrumental function width and the effect of the ion temperature on the spectral shape. However, the Stark width and thus n$_e$, can still be extracted from the wings of the spectral shape, provided a sufficient signal to noise ratio is available. To achieve this, the spectra is dynamically time-averaged over multiple frames to achieve a peak to noise level of higher than 30 for the Stark fitting. Furthermore, a weighting function is used to emphasize the importance of fitting the wings of the total line shape correctly, where the electron density is kept as a free parameter while a fixed Gaussian FWHM of a T$_i$ 3-5 eV Maxwellian ion velocity distribution [6] is assumed for the Doppler component. Comparison of the inferred Stark density is consistent across different Balmer lines fit (n=6,7,9,10,11,12,13 – note that the n=8 Balmer line is strongly polluted by a He I impurity line and hence could not be used to infer the Stark density), indicating that the influence of the Doppler line shape and any non-Maxwellian distribution [46] is negligible in the Stark density inference. The main contributors to the Stark density uncertainty are instrumental function uncertainties and, when the spectra is dynamically averaged to improve S/N ratio, an uncertainty of ~ $10^{19}$ m$^{-3}$ (or 20% - whichever is higher) is estimated. These characteristic uncertainties are obtained by comparing simplified Stark models [6, 43, 44] against more complete Stark models [46]; including uncertainties in the known electron temperature, neutral temperature, instrumental function, magnetic field as well as including realistic signal to noise ratio levels are included in a Monte Carlo fashion (section 4) when applying Stark broadening. This technique, as well as further details on Stark broadening inferences on TCV are provided in [10].

## 3.2 Separating excitation and recombination contributions of $B_{n \to 2}$

After a density estimate through Stark broadening is obtained, excitation and recombination contributions to the Balmer line emission along a viewing chord are separated quantitatively using Balmer line ratios [6], under the assumption that the neutral fraction is fairly constant (section 3.6.3), enabling the simultaneous determination of the recombination and ionisation rates. This technique has already been developed for recombination in [6] and has been optimised here to enable a more accurate excitation emission estimate, necessary for estimating the ion source ultimately. In addition, also the applicability of the technique for obtaining the ion source as well as a more thorough analysis into line-integration effects on this separation of excitation/recombinative emission is presented.

First, we provide an overview of the technique presented in [6]. For a given electron density and neutral fraction, both the line ratio between two Balmer lines ($B_{n_2 \to 2}/B_{n_1 \to 2}$) and the fraction of emission due to recombination of a certain Balmer line ($F_{rec}(n)$) become functions of only the electron temperature. This provides a relation between $B_{n_2 \to 2}/B_{n_1 \to 2}$ and $F_{rec}(n)$ over which the electron temperature varies, as shown in Figure 3 for the $B_{6 \to 2}/B_{5 \to 2}$ ratio (similar for other line ratios). A unique solution for $F_{rec}(n)$ from the measured ratio of two Balmer lines is obtained when $F_{rec}(n)$ is in between ~ 0.15 and ~ 0.85 – limiting the applicability of this technique in the low/high $F_{rec}(n)$ regions. Figure 3b is shown here for comparison against figure 3a, but will be discussed in section 3.6.3.

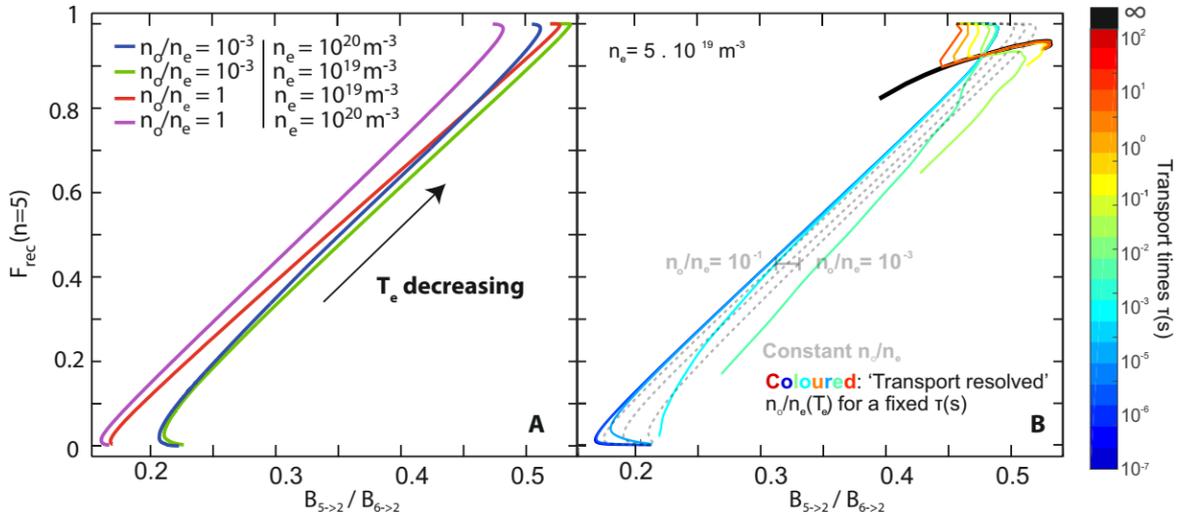

*Figure 3: $F_{rec}(n = 5)$ as function of $B_{6 \to 2}/B_{5 \to 2}$ for various electron densities and neutral fractions a). b) Same as Figure 3a, but with added transport resolved $n_o/n_e$ ($T_e$) calculation for a range of different hydrogen residence times τ [47], including a full equilibrium (τ → ∞). This will be treated in section 3.6.3.*

It is clear from Figure 3 that the $F_{rec}(n)$ obtained from the Balmer line ratio is relatively insensitive to the electron density and neutral fraction making it strongly insensitive to line integration effects, which will be further discussed in section 5. The characteristic uncertainty of $F_{rec}(n)$ is ~ ± 0.1, which was determined in this work through a probabilistic analysis presented in section 4. When determining $F_{rec}$, a single temperature is used for the excitation/recombination region for simplicity. This is consistent when comparing this with a post-processed calculation using the separate output temperatures of the excitation/recombination rate analysis [10]. In other sections of the paper a different temperature is ascribed to both regions.

New in this analysis is the realisation that the Balmer line pair used in the analysis must be chosen appropriately, especially when trying to obtain ionisation estimates, depending on the expected dominance of recombination (e.g. recombination to ionisation ratio - $R_L/I_L$) and electron density. This is illustrated in Figure 4 where, for a fixed neutral fraction, the variation of $F_{rec}$ is given for four Balmer lines as a function of temperature and density. For a fixed neutral fraction/electron density, a decreasing temperature is accompanied by an increase in $R_L/I_L$, which is also shown in Figure 4. Temperature ranges are indicated in Figure 4, because a single value for $R_L/I_L$ corresponds to a range of $T_e$ when considering the large window of electron density [$10^{18} – 10^{21}$] m$^{-3}$. When values of $F_{rec}$ (or $F_{exc}$ = 1-$F_{rec}$) are less than roughly 0.1 the uncertainties in determining the recombination (ionization) rate are too large Thus, inferring both the ionisation and recombination rate simultaneously can only be performed for a range, or window, of plasma conditions. These are different for each Balmer line; as the n level is decreased, the window in $R_L/I_L$ ($T_e$) shifts higher (lower). The regions where the line ratio can be used for ionisation estimates are shaded in red while the region applicable for recombination estimates are shaded in blue in figure 4. Note that this is a helpful 'guide' rather than an absolute number: e.g. there are ways (section 3.2) to obtain information on excitation/ recombination even if more than 90% of the Balmer line's brightness is due to recombination.

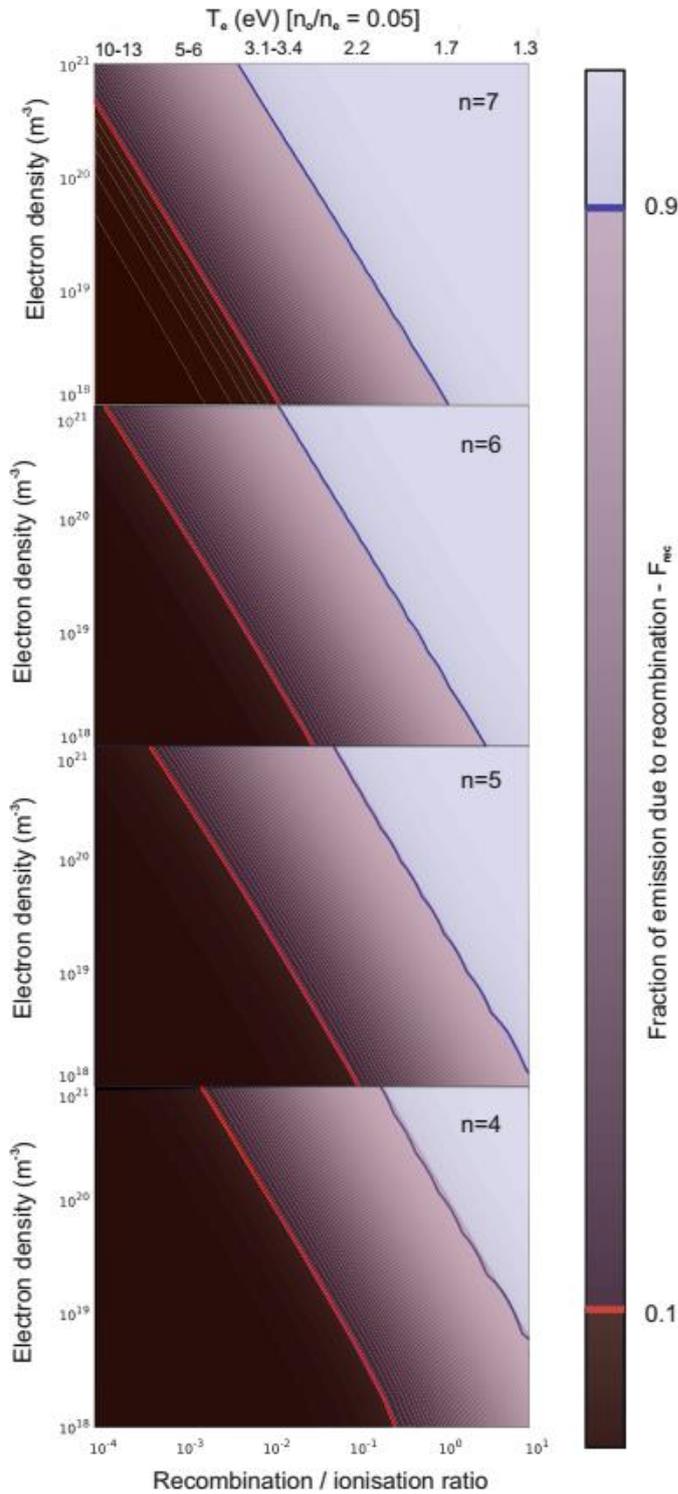

Figure 4: $F_{rec}$ as function of the electron density, ionisation to recombination ratio and Te for the n=4,5,6,7 Balmer lines, assuming $n_o / n_e$ = 0.05. Regions where $F_{rec}$>0.1 are shaded blue (recombination inference possible) and regions where $F_{rec}$<0.9 (ionisation inference possible) are shaded red, where the red and blue lines indicate $F_{rec}$ = 0.1 and $F_{rec}$ = 0.9, respectively.

The electron density, however, strongly influences $F_{rec}$ for a fixed $R_L/I_L$ as illustrated in Figure 4. This means that, as the electron density is increased, relatively more recombinative emission would occur for the same level of recombination rates. The reason for this is that three-body recombination becomes more prominent at high densities, which leads to an increased photons / recombination event ratio. This has serious implications for the applying Balmer series analysis technique to high density devices. For example, when considering C-Mod level densities of up to $10^{21}$ m$^{-3}$ about 90% of the n=4 Balmer line emission would be due to recombination even if the recombination rate is 10 times smaller than the excitation rate.

However, for typical TCV divertor conditions ($n_e$ < 5.$10^{19}$ m$^{-3}$, $R_L / I_L$ < 0.1), this means that using the n=6,7 Balmer lines suffices for extracting densities, ionisation rates and recombination rates. However, for a relatively dense TCV divertor (> $10^{20}$ m$^{-3}$), which can be achieved during L-mode density ramps with $I_p \geq$ 340 kA, $R_L/I_L$ > 5% yields $F_{exc}$ (n=6,7) < 0.1 and thus a lower-n Balmer line (n=5 or n=4), which has a larger $F_{exc}$, is required for ionisation inferences. Using the n=5 Balmer line, as opposed to higher-n Balmer lines, does mean that the sensitivity to detecting recombination becomes weaker, but recombination remains detectable when it is increases above ~2% of the ionisation rate. Since high spectral-resolution data of the n=7 Balmer line is still required for electron density estimations through Stark broadening, obtaining high spectral-resolution measurements for both the n=5,7 Balmer line with the single available TCV DSS spectrometer required discharge repeats with different spectrometer wavelength ranges.

Moving to using lower n Balmer transitions for determining the ionization rate as the divertor density increases has a limit: molecular reactions [17] will likely contribute significantly to the lowest-n Balmer line intensities (see section 3.5).

However, through our experience, we have found that one can also use physical expectations to filter unrealistic artefacts in the analysis which can appear in limiting regimes where $F_{rec}$ ~ 0.9; enabling one to apply the analysis even in cases of higher $F_{rec}$; two numerical algorithms for this are highlighted below. Those algorithms make use of the assumption that the temperature at the excitation emission region along the line of sight drops (not rises) during a density ramp/seeding scan. Such techniques are also relevant for TCV divertor conditions where $F_{rec}(n \geq 5) > 0.9$ can be reached.

As $F_{rec}$ approaches 1 the upper uncertainty bound is reduced given the limit of 1, potentially leading to an underestimation of $F_{rec}$. Naturally, if $F_{rec} > 0.9$ even small underestimates of $F_{rec}$ (n) can lead to a large overestimation of $B_{n \to 2}^{exc}$. This effect is amplified by the strong increase in the Balmer line emission (factor ~50 [6] (n=6,7)) at detachment. The combination of those effects can lead to an unphysically rapidly increasing excitation emission brightness (and thus ionisation rate – section 3.) if $F_{rec} > 0.9$ during detachment. However, that rapid increase in the excitation brightness (over time during a density/impurity ramp), assuming a fixed path length, Stark density and neutral fraction, would imply an increase in the excitation temperature ($T_e^E$ - section 3.3). By filtering out individual analysis points of the Monte Carlo (see section 4) run where $T_e^E$ goes up during a seeding/density ramp, unphysical results of a rapidly increasing ionisation source can be removed. It is important to note that only the *trend* of $T_e^E$ for each individual Monte Carlo randomization matters for this correction procedure.

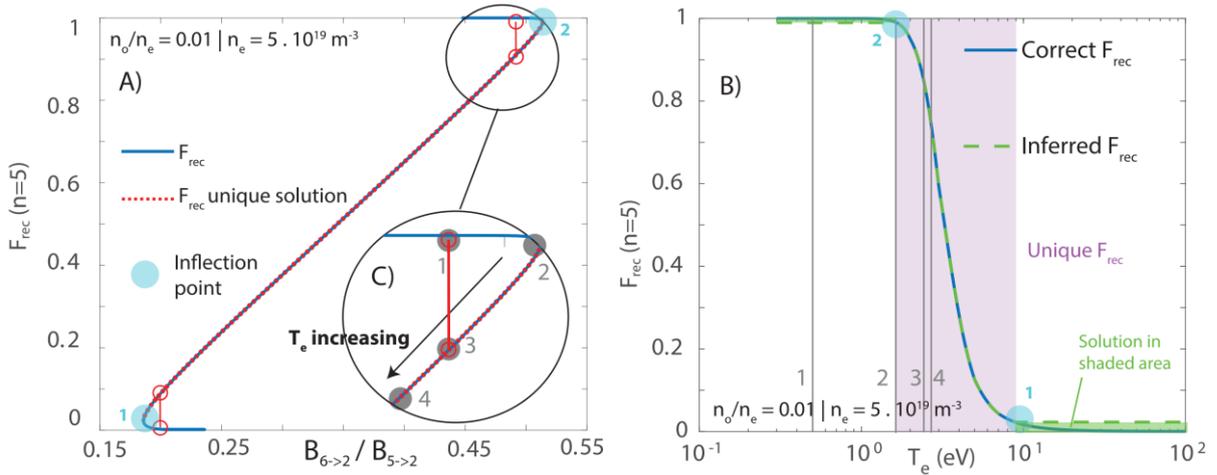

*Figure 5: An illustration of the technique for obtaining a unique $F_{rec}$: a) We first show the actual $F_{rec}$ (n=5) as function of the $B_{6\to2}/B_{5\to2}$ line ratio (labelled '$F_{rec}$'). A second version is shown ('$F_{rec}$ unique solution') which prevents non-unique solutions by limiting the used $T_e$ window. The red line with connecting circles indicates the double-valued $F_{rec}$ which is obtained from a measured line ratio which falls outside the unique solution region. b) Comparison of the inferred $F_{rec}$, given a measured line ratio, as function of $T_e$ for the 2 cases shown in Figure 5a.*

*c) (located in figure 5a) Magnification of Figure 5a in the $F_{rec}$ ~ 1 non-unique region where four points of $T_e$ are highlighted, which are represented by the numbered vertical lines in Figure 5b to link figure 5a & 5b.*

A second issue that occurs as $F_{rec}$ or $F_{exc}$ approach 1 is properly determining their values from the Balmer line ratio. As illustrated in Figure 4 with accompanying figure legend, $F_{rec}$ and $F_{exc}$, when determined from the Balmer line ratio, are double-valued (e.g. not unique) in certain ranges of $F_{rec}$: two different $T_e$ and $F_{rec}$ correspond to the same line ratio as shown in Figure 5a and b. Figure 5c provides a magnification of the $F_{rec}$ ~ 1 region in Figure 5a. Four points are illustrated in Figure 5c corresponding to four values of $T_e$ highlighted by vertical lines in Figure 5b for the reader to help connect Figure 5a to Figure 5b. As the Balmer line ratio (for fixed density and neutral fraction) changes as function of dropping temperature, it transitions from being dominated by excitation (large $F_{exc}$, low $F_{rec}$) to being dominated by recombination (large $F_{rec}$, low $F_{exc}$). At the extremes of this temperature and $F_{rec}$ range there is a loss of uniqueness in determining $F_{rec}$ and $F_{exc}$. Two inflection points of the Balmer line ratio (e.g. a minimum and a maximum) can be identified, in between which a unique solution and value for $F_{rec}$ can be obtained as indicated in figure 5a. If the temperature is dropping, one goes chronologically through three phases:

1) Before reaching the nr. 1 inflection point (figure 5), $F_{rec}$ is between 0 and the value corresponding to the inflection point. In the analysis code it is thus randomly chosen in this range.
2) Between the nr. 1 and nr. 2 inflection points the obtained value of $F_{rec}$ is unique;
3) After reaching the nr. 2 inflection point (figure 5), $F_{rec}$ is between the value corresponding to the inflection point and 1. In the analysis code it is thus randomly chosen in this range.

This is performed for every Monte Carlo randomisation and every line of sight separately. The two inflection points are obtained experimentally from the $F_{rec}$ inference. See [10] for more details. Also note that these techniques only need to be applied if the $F_{rec}$ reaches close to 1, which is the case for the discharge discussed in this work (#56567).

## 3.3 Inferring ionisation and recombination rates

Once $B^{rec}_{n \rightarrow 2}$ is obtained, the recombination rate integrated along a spectroscopic line of sight, $R_L$ [rec/m$^2$ s] can be obtained from the inferred Stark density and the assumed path length $\Delta L$ as highlighted in [6] and as shown in Figure 6a. Using an analogous approach, once $B^{exc}_{n \rightarrow 2}$ is obtained, the ionisation rate integrated along a spectroscopic line of sight [ion/m$^2$ s] can be obtained from the inferred Stark density and an estimate of the combined parameter, $\Delta L n_o / n_e$ (which represents a path length scaled neutral concentration in m), as illustrated in Figure 6b. The recombination and ionisation rates used are modelled using the so called effective recombination coefficients (ACD), and effective ionisation coefficients (SCD) from the Open-ADAS tables [40-42], which are functions of electron density and temperature. Figures 6a and b show $I_L$ is considerably more sensitive to its defining input parameters than $R_L$. It is important to note that uncertainties of both the combined parameter $\Delta L n_o / n_e$ (which has an order of magnitude larger uncertainty than just $\Delta L$ on which the $R_L$ determination depends) and $n_e$ play a major influence in the uncertainty of $I_L$. This is another reason why we use a probabilistic analysis to robustly provide an estimate for both $I_L$ and its uncertainty.

Line integration effects (section 5) lead to differences between the inferred ionisation rate and the 'true' ionisation rate. Although these differences are larger for ionisation rate than for inferred recombination rates, for both cases the uncertainty introduced by line integration effects remains smaller than the characteristic uncertainty of the quantities themselves. To minimize the uncertainty in $I_L$, the lowest n Balmer line used for determining $F_{rec}(n)$ is used to determine $I_L$, whereas the highest-n Balmer line is used to determine $R_L$. The results agree within uncertainty when either line in the Balmer line pair used to determine $F_{rec}$ (n) is used to determine $R_L$ and/or $I_L$. The results also agree when other (appropriate) Balmer lines are used which were not used to determine $F_{rec}$.

This above process of determining ionisation and recombination rates leads to an estimate of the 'characteristic' temperature of the excitation ($T_e^E$) and recombination ($T_e^R$) regions along the chordal path length. Considering how $I_L$ and $R_L$ are determined, these temperatures are defined as the necessary temperature required for the 0D plasma slab model of fixed quantities ($\Delta L, n_e, n_o$) to match the experimental $B_{exc}^{n\rightarrow 2}$ and $B_{rec}^{n\rightarrow 2}$. It has been verified that this way of obtaining $T_e^R$ yields similar results to the temperature obtained by fitting the n>9 Balmer lines with a Saha-Boltzmann functional dependence on $T_e^R$ (e.g., see tokamak applications [9, 16, 17]). Physically, $T_e^E$ and $T_e^R$ approximate a 'chord-averaged' temperature, weighted by the excitation and recombination emissivities respectively, along the line of sight.

The inference of both excitation and recombination rate temperatures is achieved through separate analyses and are usually different. $T_e^E$ inferred from our measurements ranges from ~ 3 eV to ~ 20 eV, whereas $T_e^R$ ranges from ~0.5 to ~4 eV. $T_e^E$, measured near the target, has been verified against other temperature measures, based on power balance and analytic models, in [10, 11]. This temperature difference is due to the occurrence of recombination and excitation in different locations as confirmed with SOLPS-Eirene modelling (section 5); excitation primarily occurs in higher temperature regions than recombination. Separating the excitation and recombinative emission regions, using $F_{rec}$, thus makes the analysis more robust to line integration effects. In addition, making $T_e^E$ and $T_e^R$ two separate entities enables (partially) compensating line integration effects – as the temperature at the excitation/recombination region is indeed different, making the analysis even more robust to profile related effects (section 5).

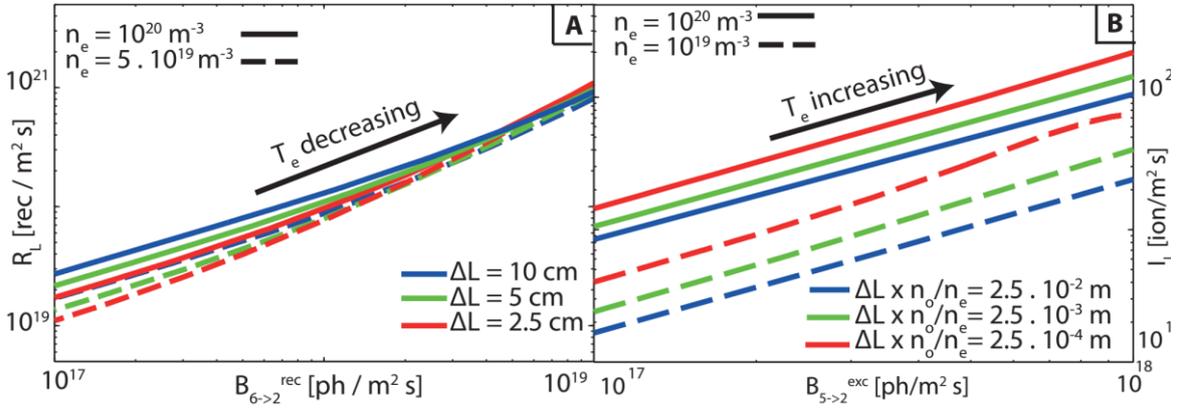

Figure 6: a) The recombination rate $R_L$ along a spectroscopic line of sight as a function of $B_{6\rightarrow 2}^{rec}$ for various $n_e$ and $\Delta L$. b) The ionisation rate $I_L$ along a spectroscopic line of sight as function of $B_{5\rightarrow 2}^{exc}$ for various $n_e$ and $\Delta L \times n_o/n_e$.

Since $T_e^E$ is obtained when obtaining a $I_L$ inference, it is possible to use this temperature, together with the Stark density to estimate the charge exchange rate to ionisation rate ratio using ADAS values [40-42] – assuming that charge exchange and excitation occur at the same location of the line integral. That is a reasonable assumption as the charge exchange rate is relatively temperature independent and the ionisation/charge exchange rates are both linearly dependent on the neutral density.

### 3.4 Estimating hydrogenic radiative power losses

The amount of energy expended per ionization is central to our analysis of the role of power balance in the ionization process. To obtain this, we obtain first the total power loss due to ionisation over a chordal integral, $P_{Ion,L}$ (W/m$^2$). This is modelled in Eq. 7a using the 0D plasma slab model with spatially constant parameters as function of $T_e^E$, $n_o$, $\Delta L$ and $n_e$. The first part of $P_{Ion,L}$ represents the radiative losses associated with ionisation ($P_{rad,L}^{exc}$), which occur during the several, to multiple, excitation collisions where energy is lost due to line radiation as the atom deexcites before finally ionizing; which is modelled using the ADAS PLT coefficient (in W m$^3$), defined as the excitation-related radiative power loss rate. This is obtained within ADAS by integrating over the full modelled excitation spectra based on $n_e$, $T_e$ [40-42]. The second contribution to $P_{ion,L}$ is the energy removed from the plasma and stored as the potential energy of a new ion, corresponding to $\epsilon$ = 13.6 eV [1, 47], where molecular dissociation is ignored, which is obtained by multiplying the ionisation rate $I_L$ with $\epsilon$, resulting in $P_{ion,L}^{pot}$ in Eq 5a.

One important physical parameter is the energy 'cost' per ionisation (averaged along a chord, weighted by the ionisation profile along the chord) [11, 12], which is obtained by dividing $P_{ion,L}$ with $I_L$, resulting in equation 7b. Detailed discussions on the importance of this parameter and experimental measurements using this technique can be found in [11], which shows that the radiated energy per single ionisation event (and thus $E_{ion}$) increases strongly at low electron temperatures (see, for example, [38, 48, 49]). These TCV measurements indicated that ~ 25 eV per ionisation is needed during the attached phases of the analysed discharges, increasing to above 80 eV in colder regions below the peak ionisation region during detachment (~40 eV averaged over the divertor).

The experimental estimation of the power loss due to ionisation is analogous to the ionisation rate inference in section 3.3 (Figure 6b); by using the excitation brightness, Stark density and by using an estimate of $\Delta L \times n_o/n_e$. The results, shown in Figure 7, are identical to using Eq. 7a in combination with the temperature of the excitation region obtained previously during the ionisation rate inference.

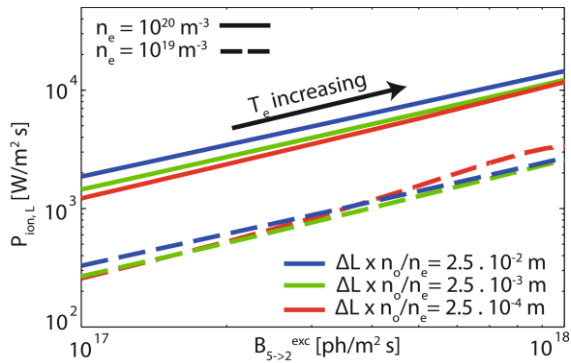

Figure 7: Power required for ionisation along a line of sight as function of the emission due to excitation (n->5) for various levels of electron density (solid vs dashed) and $\Delta L \times n_o/n_e$.

$$P_{ion,L} = \underbrace{\Delta L\, n_e\, n_o PLT(n_e, T_e)}_{P_{rad,L}^{exc}} + \underbrace{\Delta L\, n_e\, n_o\, \epsilon SCD(n_e, T_e)}_{P_{ion,L}^{pot}} \quad (7a)$$

$$E_{ion} = \frac{PLT(n_e, T_e)}{SCD(n_e, T_e)} + \epsilon \quad (7b)$$

Using a similar approach, the Open-ADAS PRB parameter [40-42] (in W m$^3$), defined as the recombination (radiative and three-body) and Bremsstrahlung related radiative power loss rate, can be used to estimate the radiated energy losses due to recombination per recombination reaction. The PRB parameter combines both radiated power due to recombination (three-body and two-body recombination; including both line emission and continuum emission) and bremsstrahlung (electron - hydrogen ion collisions). Since we are only interested in radiative losses due to recombination, the Bremsstrahlung component (whose contribution is negligible for recombination-relevant temperatures) is subtracted from the PRB coefficient by modelling the radiated power due to Bremsstrahlung as listed in [50]. The radiative losses due to recombination are found to be 12.5-14.5 eV per recombination reaction (see [11]), which are very similar to $\epsilon$: for TCV-relevant densities the radiative energy loss during volumetric recombination and the potential energy gain thus seem to approximately cancel.

## 3.5 The influence of molecules on this analysis and the 'up-/down-conversion' of Balmer line emission

Separating the Balmer line emission quantitatively into excitation/recombination contributions, and obtaining characteristic densities/temperatures for both regions, provides one with all the information in the *atomic* part of the Balmer line series. This implies that, having quantitative numbers of the excitation/recombination emission of any Balmer line, together with their respective temperatures and density (here assumed to be equal – e.g. the Stark density), enables one to predict the entire *atomic* hydrogen spectra (in theory also the Lyman/Paschen series could be modelled through this). If $T_e^E$, $T_e^R$, $B_{n\rightarrow 2}^{exc}$, $B_{n\rightarrow 2}^{rec}$ of a Balmer line $n_1$ is known (which is all provided by the analysis steps above), the resulting emission of a Balmer line $n_2$ is given by:

$$B_{n_2\rightarrow 2} = B_{n_1\rightarrow 2}^{rec} \times \frac{PEC_{n_2\rightarrow 2}^{rec}(n_e, T_e^R)}{PEC_{n_1\rightarrow 2}^{rec}(n_e, T_e^R)} + B_{n_1\rightarrow 2}^{exc} \times \frac{PEC_{n_2\rightarrow 2}^{exc}(n_e, T_e^E)}{PEC_{n_1\rightarrow 2}^{exc}(n_e, T_e^E)} \qquad (8)$$

Using this technique has enabled us to predict various Balmer line intensities within uncertainty, except for the $D_\alpha$ brightness. This is used in [10] to indicate a significant contribution (of more than 50%) of molecules to the $D_\alpha$ intensity during detached TCV density ramp discharges – in quantitative agreement with SOLPS simulations, which will be further investigated in the future.

The presence of plasma-molecule interactions can influence the analysis in this paper in two ways. First, plasma-molecule interactions can serve as additional power/particle sinks and sources, in addition to the *atomic* particle sinks/sources inferred in our analysis. We do not currently have a technique, of relatively similar difficulty to that described above for atomic processes, to easily determine the molecular-activated recombination (MAR) and ionization (MAI) rates which may be significant. However, the observed Dα and accompanying analysis during a density ramp on TCV may be indicative of the presence of MAR during detachment [10]. Secondly, plasma-molecule interaction could contribute to the Balmer line emission, especially to $D_\alpha$ [14, 51]. This contribution is thought to be reduced/negligible for n>4 Balmer lines [52-56]. This limits the lowest-n Balmer line usable in the described analysis and we only apply it to n=5 or higher Balmer lines.

## 3.6 Input parameters and their uncertainties

A summary of the required input parameters for the analysis is shown in table 1 and figure 2. As explained in the introduction, a probabilistic Monte Carlo approach has been used to propagate the uncertainties fully into the result and thus a probability density function for each input parameter must be assumed (see section 4 for more details). An estimate of such uncertainties is shown in table 1.

The input parameters required include directly measured parameters, such as the Balmer line intensity, ratio between two Balmer lines and the Balmer line shape leading to a Stark density. The Stark density inference is covered in section 3.1. The uncertainties in the brightness and Balmer line ratio originates mostly from (absolute/relative) calibration uncertainties [6], which are estimated to be significant due to the calibration of the system in the near-UV [10].

Two other input variables are required which are estimated, not measured: the path length estimate $\Delta L$, which is an estimate of the characteristic length of the region where most of the Balmer line emission occurs along the chordal integral, and an estimate of the neutral fraction $n_o/n_e$, which is the ratio between hydrogen neutral and electron density (which is assumed to be equal to the electron density. That assumption (e.g. assuming $Z_{eff}$ = 1) is expected to have a negligible influence on the presented analysis based on initial testing in [6]. Those two non-directly measured input parameters ($\Delta L$ and $n_o/n_e$) are discussed in more detail in the two subsequent subsections.

| Name | Parameter | Unit | Uncertainty / probability density function |
|---|---|---|---|
| Balmer line brightness (n) | $B_{n\to 2}$ | ph/m² s | Gaussian<br>Peak: Measured brightness<br>68% conf. interval: 17.5% |
| Balmer line ratio ($n_1$, $n_2$) | $\dfrac{B_{n_1 \to 2}}{B_{n_2 \to 2}}$ | - | Gaussian<br>Peak: Measured line ratio<br>68% conf. interval: 12.5% |
| Path length | $\Delta L$ | m | Asymmetric Gaussian<br>Peak: Inferred (see 3.6.1)<br>Upper uncertainty: 50%<br>Lower uncertainty: 20% |
| Neutral fraction | $n_o/n_e$ | - | Uniform (or log-uniform)<br>$10^{-3}$ – 0.05 |
| Stark density | $n_e$ | m⁻³ | Gaussian (with minimum cut-off)<br>Peak: Inferred density from fit<br>68% conf. interval: 20% or $10^{19}$ m⁻³<br>Minimum cut-off: [0.1-0.5] · $10^{19}$ m⁻³ |

*Table 1: Overview of the various analysis inputs, together with their estimated uncertainty/probability density function.*

### 3.6.1 Path length estimates

As described above, an estimate of the path length, $\Delta L$, through the region of strongest contribution to the measured brightnesses, is required. We estimate $\Delta L$ (at the target) to correspond to the full-width-half-maximum of the ion target flux profile measured by Langmuir probes and its two corresponding flux surfaces (see Figure 8a). The $\Delta L$ for other points along the divertor leg is then calculated as the distance between the intersection of the spectroscopic lines of sight with these mapped flux surfaces (see Figure 8a). Since the $J_{sat}$ SOL width increases during a core density ramp [57], the defined $\Delta L$ is determined as function of time (Figure 8b). Comparing the estimated value of $\Delta L$ using a synthetic diagnostic with the Balmer line emission profile along the line of sight obtained from SOLPS simulations (section 5), shows that the estimated $\Delta L$ corresponds to a region where at least 70% of the Balmer line emission occurs [10]. In future work, the $\Delta L$ estimate could be improved utilising filtered camera imaging [58].

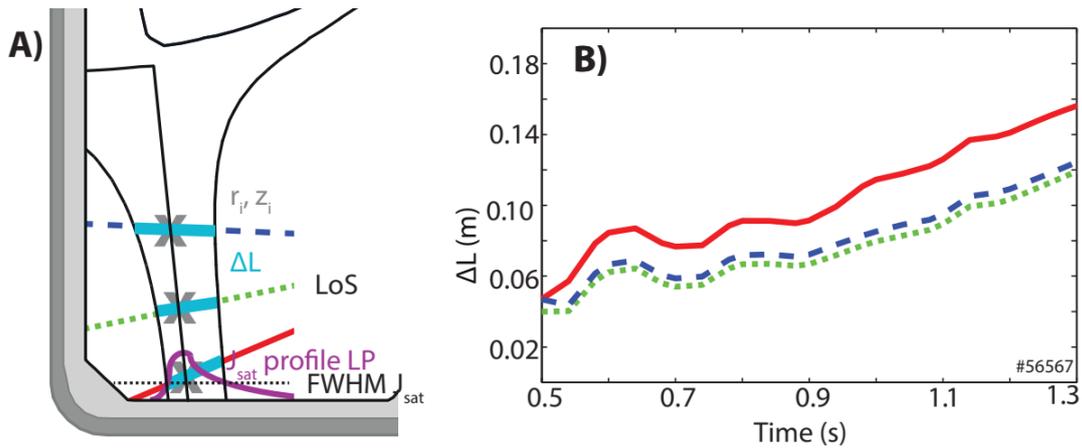

*Figure 8: a) Cartoon illustrating how pathlength ($\Delta L$) is determined using the FWHM of the $J_{sat}$ profile, together with the profile locations ($r_i$, $z_i$); which are the intersections of the lines of sight with the separatrix (X); of DSS inferences along the outer divertor leg. b) Example of $\Delta L$ as function of time for three lines of sight.*

### 3.6.2 Neutral fraction

To perform the analysis illustrated, a neutral fraction – e.g. the ratio between the neutral hydrogen density and the hydrogen ion density – must be assumed. That ratio is used when separating the excitation/recombination contributions to Balmer line emission and for the excitation emission calculations. The neutral fraction is not well known experimentally and thus we assume that the neutral fraction is somewhere within a range, covering over more than an order of magnitude, obtained from various modelling including both new [27] and older [29, 59] TCV SOLPS simulations, as well as OSM-Eirene interpretive modelling [7].

Although separating excitation/recombination emission does not depend strongly on the neutral fraction (section 3.2 figure 3); it does depend on the assumption that the neutral fraction – at the excitation emission region – is 'fairly' independent of temperature: e.g. it is a fixed constant rather than a function of temperature. Such a function of temperature would, for instance, be obtained when assuming there is no transport and a local equilibrium between excitation/recombination exists – for more information see section 3.6.3. This assumption implies that the neutral transport in the divertor is such that the influence of the creation of neutrals (volumetric recombination) and destruction of neutrals (ionisation) - two processes which change critically as function of temperature – is negligible on the actual neutral density at the excitation emission region. Under this assumption, as the temperature drops, a transitioning of the Balmer line ratio is expected indicating a transitioning of excitation dominant Balmer line emission to recombination dominant is expected; as indicated figure 9. If, instead, one was to assume that neutral transport can be neglected [34, 37] (e.g. the neutral fraction is determined due to a local balance of ionisation and recombination – calculated using effective ADAS ionisation/recombination rates – see section 3.3) – a completely different behaviour of the Balmer line ratio, as indicated in figure 9. In other words, the trend of the Balmer line ratio depends crucially on the neutral dynamics in the divertor.

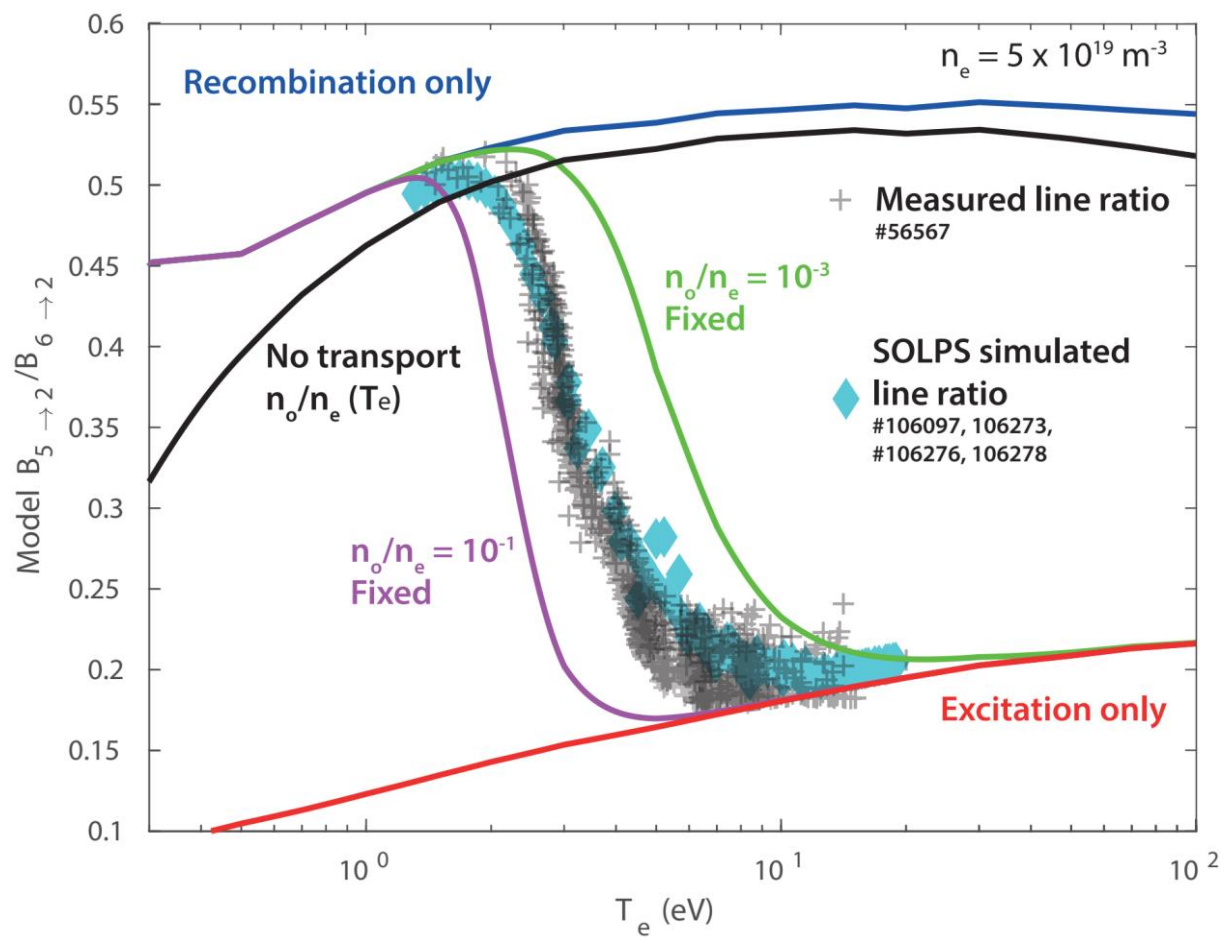

Figure 9: n=5/n=6 Balmer line ratio shown as function of electron temperature. The Balmer line ratio is modelled using Open-ADAS showing two separate trends for excitation only (lower line ratio) and recombination only *(higher line ratio) emission. Assuming a fixed neutral fraction $n_o/n_e$ leads to a modelled 'jump' in the Balmer line ratio, reminiscent of a transitioning of excitation to recombination Balmer line emission. Assuming a local ionisation/recombination balance equilibrium (labelled 'no transport'), as done in [34, 37], yields a different trend. The measured Balmer line ratio for all spectrometer lines of sight during a density ramp is also shown as function of the excitation derived temperature, as well as a Balmer line ratio obtained from SOLPS simulations using a synthetic diagnostic approach as function of the excitation emission weighted temperature along each spectroscopic chord.*

This assumption is verified using recent TCV SOLPS simulations [27] in [10] where - using a synthetic diagnostic technique - the *excitation emission weighted average* $n_o/n_e$ along each DSS line of sight is between 0.01 – 0.035 during an upstream density scan, where the plasma is first attached, later detached and ultimately deeper detached (higher upstream density) than the experiment. However, modelling the neutral fraction as a local balance of ionisation and recombination would lead to a range of neutral fractions of $10^{-6}$ to $10^4$ [10]. In other words, the simulations indicate that the neutral fraction is affected by transport in such a strong way that the neutral fraction is relatively constant.

The measured Balmer line ratio during the experiment also indicates a smooth transitioning from excitation dominated emission to recombination dominated emission, which is observed as function of time during a density ramp experiment where the divertor temperature is continuously decreased [10, 11]. This is visualised in figure 9 using the measured n=5/n=6 Balmer line ratio as function of $T_e^E$. For completeness, the same Balmer line ratio obtained from the DSS chords using a synthetic diagnostic on the SOLPS simulations representing this density ramp has been shown as function of the excitation-emission weighted temperature along the DSS lines of sight; indicating the same trend and magnitude as observed experimentally.

The above investigation clearly indicates that the neutral dynamics in TCV are driven by transport and can be assumed to be relatively insensitive of local temperature – in quantitative agreement with SOLPS simulations.

### 3.6.3 Influence of a $n_o/n_e$ ($T_e$) dependency on $F_{rec}$

Although we have shown above, using Balmer line ratio trends, that the neutral dynamics in TCV is so strongly driven by transport that it can therefore be assumed to be quasi-constant, we can investigate how a $n_o/n_e$ ($T_e$) influence changes the relation between $F_{rec}$ and the Balmer line ratio – which could be relevant for applying the analysis technique developed in this work to other devices. For that it is important to realise that, actually (e.g. more accurately), the assumption made is that any $n_o/n_e$ ($T_e$) influence does not change the picture/conclusion of figure 3a (e.g. the relation between $F_{rec}$ and the Balmer line ratio) significantly. To investigate this, we utilise a transport-resolved model for $n_o/n_e$ ($T_e$, $\tau$), where $\tau$ is an assumed residence time required for a specie to establish equilibrium [36, 42, 60, 61] using the Open-ADAS effective ionisation/recombination rates [40-42], which is strongly correlated with the ionisation state distribution at equilibrium. For $\tau \rightarrow \infty$, the result converges to the no-transport result.

The result, shown for a large range of $\tau$ as well as the 'equilibrium' case ($\tau \rightarrow \infty$) in Figure 3b, indicates that a transport-resolved $n_o/n_e$ ($T_e$) behaviour has a very similar relation between $F_{rec}$ and the Balmer line ratio than assuming $n_o/n_e$ is a constant. The main difference between a $n_o/n_e$ ($T_e$) and a fixed $n_o/n_e$ occurs at high values of $F_{rec}$, where transport can lead to a limited regime in which $F_{rec}$ is reduced from ~1 to ~0.9 at high recombinative regimes as the recombination process generates neutrals – which in turn can lead to excitation emission if a significant amount of neutrals is generated (e.g. neutral fractions > 1; which is, at least, significantly higher than obtained from TCV divertor modelling (section 3.6.2) and may be unphysical also in other divertors). Such a strong influence of recombination on the neutral fraction may lead to a small underestimation of $F_{rec}$. Therefore, the highlighted technique should be applicable even in cases where recombination/ionisation has a strong influence of the neutral fraction.

## 3.7 Summary of output parameters

The different output parameters, summarised in table 2, can be classified in two different types: chordal integrated parameters – e.g. recombination rate $R_L$ (in rec./m² s); ionisation rate $I_L$ (in ion./m² s); charge exchange to ionisation ratios $CX_L / I_L$ (unitless) and hydrogenic radiation rates (excitation $P_{rad,L}^{H,exc}$) /recombination $P_{rad,L}^{H,rec}$ in W/m²) and local parameters, such as the Stark density (in m$^{-3}$) and electron temperature (in eV). The spectroscopic coverage of the TCV divertor spectrometer leads to a full profile of those parameters along the outer divertor leg (see figure 1). By mapping the line integrated values (e.g. /m²) to single $R_i$, $z_i$ locations (prescribed to the intersection of the line of sight and the separatrix), they can be integrated toroidally (e.g. $2 \pi R f(R) dR$) and then poloidally along the path of $R_i$, $z_i$, ultimately providing total estimates of the outer divertor recombination $I_r$ (in rec./s), ionisation $I_i$ (in ion./s) and hydrogenic radiation (excitation $P_{rad}^{H,exc}$/recombination $P_{rad}^{H,rec}$ in W) – similar to the approach used in [17] for the recombination rate. All output parameters, except the Stark density, have a probabilistic uncertainty analysis (section 4). The inference of all these output parameters is investigated using synthetic diagnostic routines applied to validated SOLPS simulations in section 5.

| Type | Name | Parameter | Unit | (R)ecomb./(E)xcit. |
|---|---|---|---|---|
| Tor. ∫ | Ion source | $I_L$ | ion./s | E |
| Tor. ∫ | Radiated power, excit. | $P_{rad}^{H,exc}$ | W | E |
| Tor. ∫ | Recombination sink | $I_r$ | ion./s | R |
| Tor. ∫ | Radiated power, recomb. | $P_{rad}^{H,rec}$ | W | R |
| Line ∫ | Ion source | $I_L$ | ion./m² s | E |
| Line ∫ | Radiated power, excit. | $P_{rad,L}^{H,exc}$ | W/m² | E |
| Line ∫ | Charge exchange to ionisation ratio | $CX_L/I_L$ | - | E |
| Line ∫ | Recombination sink | $R_L$ | ion./m² s | R |
| Line ∫ | Radiated power, recomb. | $P_{rad,L}^{H,rec}$ | W/m² | R |
| Line ∫ | Balmer line recombination emission fraction | $F_{rec}(n)$ | - | E & R |
| Line ∫ | Balmer line excitation emission fraction | $F_{exc}(n)$ | - | E & R |
| Local | Excitation temperature | $T_e^E$ | eV | E |
| Local | Recombination temperature | $T_e^R$ | eV | R |
| Local | Stark density | $n_e$ | m$^{-3}$ | E & R |

*Table 2: Overview of the various analysis outputs. 'Tor. ∫' implies toroidal integral while 'Line ∫' implies chordal integral. E and/or R indicates whether the parameter was derived from recombination or excitation emission.*

## 4 Probabilistic Monte-Carlo analysis

Given the complicated set of analyses together with multiple input parameters shown in Figure 2, we have developed a Monte Carlo based probabilistic analysis to more accurately characterize both the most probable values of our analysis outputs (recombination and ionization rates, hydrogenic radiative losses, charge exchange rates) as well as their uncertainties in the form of probability density functions (PDFs). This analysis also makes the result less prone to errors in the input parameters. An important part of this process is that the functional form for the uncertainty of each input parameter must first be properly characterized as a PDF – ranging from Gaussian (e.g. $B_{n \rightarrow 2}$) to asymmetric Gaussian (e.g. $\Delta L$ ) to flat (e.g. $n_o/n_e$) as summarised in table 1. According to those input PDFs, random values of each input parameters are obtained through rejection sampling. Those randomisations are kept the same for all lines of sight and time steps. This means that, for instance, if the randomly sampled input values correspond to a brightness 10% below the measured brightness, this 10% is used for all time steps and all lines of sight. Keeping the randomisation the same enables utilising the techniques from section 3.2 to resolve limit conditions ($F_{rec}>0.9$), which are employed *to*

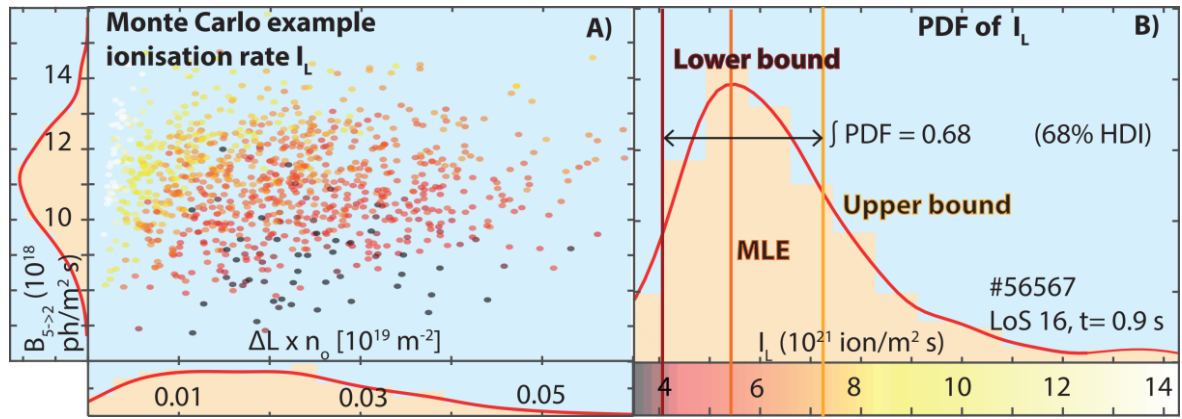

*Figure 10: Example of probabilistic analysis for one time point which shows in a) a scatter plot of the randomly chosen values for the brightness and $\Delta L \times n_o$, whose colour coding corresponds to the value of the ionisation rate shown below b) the PDF of the ionisation rate, together with the estimate of the parameter (Maximum Likelihood Estimate (MLE)) and with its 68% Highest Density uncertainty Interval (HDI).*

*each individual (at least 5000) randomisation.* Toroidally and poloidally integrated quantities are obtained for each randomization separately. Keeping the randomisation the same for all time steps is also a more realistic description of the uncertainty as the dominant uncertainties in the analysis are of a systematic nature (e.g. such as calibration uncertainties), rather than random noise.

An example of the probabilistic analysis is shown in Figure 10 for the line integrated ionisation rate $I_L$ (result for a single chord, at a given time), which – in this case – is mostly influenced by uncertainties in $B_{5\rightarrow 2}$ as well as $\Delta L \times n_o/n_e$. The scatter plot of Figure 10a (one time point) thus shows the randomly sampled values of the distributions of $\Delta L \times n_o/n_e$ and $B_{5\rightarrow 2}$ from their uncertainty PDFs which are shown as histograms to the sides of Figure 10a. A 'kernel density estimate' (a statistical non-parametric technique for providing smooth estimates for probability density functions) is employed to convert the analysis outputs into a PDF using an adaptive kernel density estimation algorithm [62]. The colour of each point in the scatter plot (Figure 10a) corresponds to an ionization rate given in the colour bar below the resultant PDF of the ionisation rate of Figure 10b.

We apply analysis techniques adopted from Bayesian analysis [63] to extract information from the PDFs. The uncertainty of the estimate is given by the shortest interval whose integral corresponds to the requested uncertainty range; commonly referred to in literature [64] as the "Highest Density Interval (HDI)"; which provide the upper and lower uncertainties for our estimates. For unimodal PDFs, this interval also contains the Maximum Likelihood Estimation (MLE) (peak) of the PDF, which we use as an estimate for the resulting parameter since it has the highest probability to occur. There are also other techniques to extract information from PDF. For example, we have compared the above techniques to taking the median value for the estimate and the 'equal-tail' [64] interval (68% probability) for the uncertainty and found essentially the same result.

The uncertainty margin for $I_L$ can be strongly asymmetric as shown by the asymmetric tail of the PDF in Figure 10b. Based on a comparison of this high $I_L$ asymmetric tail to the Monte Carlo scatter plot result of Figure 10a, we can conclude that low values for the neutral fraction lead to higher ionisation rates. Lower neutral fraction corresponds to higher $T_e^E$ which is needed to match the measured excitation emission as the number of ionisations per emitted excitation photons increases at higher temperatures.

Figure 11 provides a more general overview of the characteristic output PDFs, shown for 3 timesteps during a density ramp discharge (#56567) – respective of attached (~0.5 s), detachment-onset (~1.0 s) and detached (~1.2 s) operation. That particular discharge is discussed in more detail in section 6 and [11]. Over these three different times, the magnitude of the uncertainty and its asymmetry can vary strongly. In addition, as the fraction of emission due to recombination changes for a particular chord, the input parameters contributing most to the resulting uncertainty can change strongly; as has been investigated using a Kendall rank correlation technique in [11]. In other words: the main uncertainty cannot be attributed to a single (set of) parameters and depends strongly on the conditions present. Furthermore, it motivates the Monte Carlo approach further as this approach can capture all uncertainties in a realistic manner. Despite all these variations, the PDF of the output parameters in general remains unimodal as shown in figure 11, except for $F_{rec}$ (most likely due to the techniques highlighted in section 3.2) – which is important for the interpretation of the result. However, $F_{rec}$ is only an intermediate result.

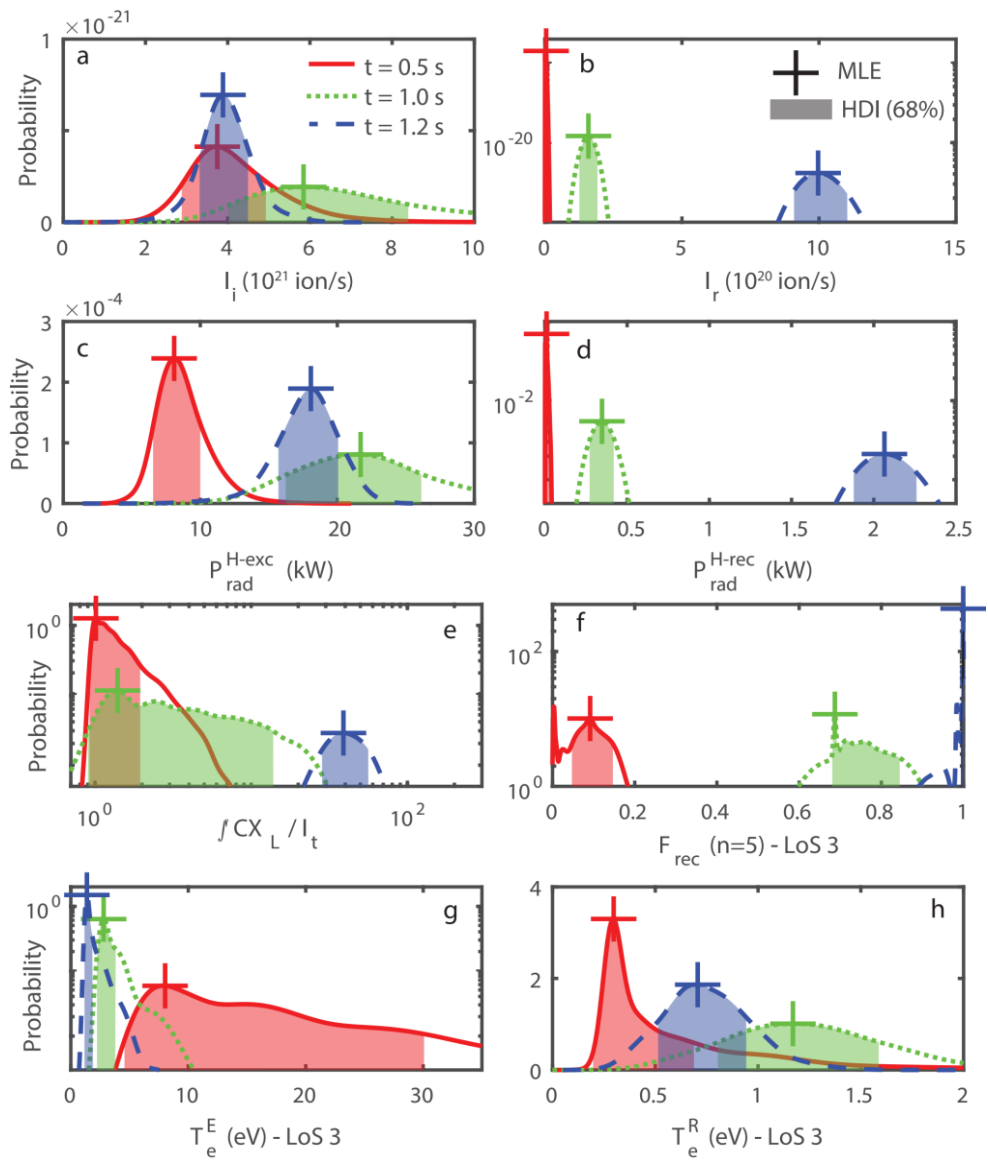

Figure 11: Characteristic Probability Density Functions from DSS output parameters ($I_i$, $I_r$, $P_{rad}^{H-exc}$, $P_{rad}^{H-rec}$, $F_{rec}$ (n=5), $T_e^E$ and $T_e^R$) taken from the analysis of #56567 at three time points corresponding to attached (~0.5 s, red), detachment-onset (~1.0 s, green) and detached (~1.2 s, blue) operation, together with the maximum likelihood estimate (MLE) and the highest density interval (HDI – 68%) corresponding to the 1 σ confidence interval. The *results shown in figure a-e correspond to outer-divertor integrated results while the results shown in f-h (e.g. ($F_{rec}$ (n=5), $T_e^E$ and $T_e^R$)) correspond to inferences from the third chord closest to the target. The results are shown at t=0.5, 1.0 and 1.2 s.*

It should also be noted that the output results will depend on the assumed input PDFs. For instance, if instead of a uniform distribution for $n_o$ (each value within the range has equal probability) a log-uniform distribution for $n_o$ is assumed (e.g. the probability of a value being between $10^{-3} – 10^{-2}$ is the same as the probability of a value being between $10^{-2} – 10^{-1}$), the resultant ionisation rate increases by ~ 20% as shown in [10].

# 5. Verification of analysis technique against SOLPS simulation solutions using a synthetic diagnostic approach

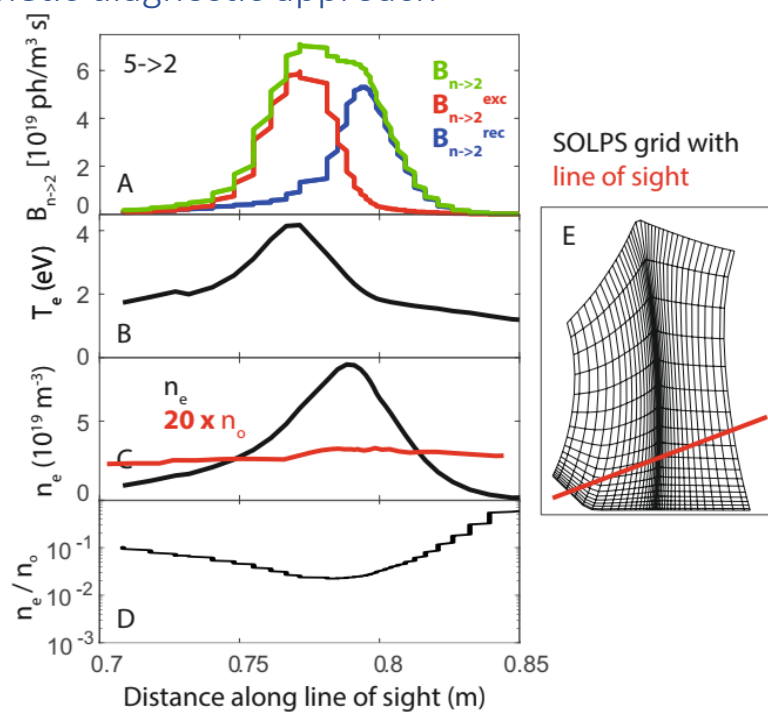

*Figure 12: SOLPS (#106273) simulated Balmer line emission profile (a) along lower line of sight (e), together with the electron temperature (b), electron, neutral density (c) and $n_o/n_e$ profiles (d). The emission profile of excitation/recombination is identical (in shape) between different Balmer lines; only the fraction of how each profile contributes to the total emission profile varies between different Balmer lines.*

The analysis highlighted in this work for the experimental determination of local (density/temperature) and line-integrated quantities (ionisation rates/recombination rates/charge exchange rates/radiated hydrogenic power) utilises a 0D plasma slab model. However, in reality, the lines of sight intersect regions of the plasma with varying temperature, electron density and neutral density; which could influence the Balmer line analysis significantly – a common drawback of passive emission spectroscopy [56]. This is illustrated in Figure 12 where the SOLPS-obtained electron density, electron temperature, neutral density and excitation/recombination emission profile is shown along the line of sight for a particular line of sight. As is shown, the temperature and density is significantly different in the emission regions of excitation/recombination; further motivating that the excitation and recombination region through passive line of sight spectroscopy cannot be described with a single electron temperature. In addition to these line integration effects, various other assumptions (highlighted in section 3) were employed in the analysis, regarding the neutral fraction, path length and plasma purity ($Z_{eff}$ = 1).

In this section we investigate the sensitivity of the techniques presented in section 3 to these assumptions by applying the analysis technique described above also to synthetic diagnostic data obtained from plasma solutions determined through SOLPS-ITER [27] – which includes the Eirene Monte Carlo model of neutral transport. The five SOLPS simulations (stored as MDS+ data according to the shot numbers noted in figure 13) used mimic a TCV L-mode density ramp [27] and include chemical sputtering of the carbon tiles to reach realistic carbon concentrations in the divertor (which have been verified against the experiment using the absolute CIII (465 nm) brightness). The upstream density in the simulations is reached through an upstream gas puff [27].

5.1 Methodology

The Balmer line emission and line shape is modelled, using Open-ADAS [40-42] tables identical to the experimental analysis, at every grid cell of the simulation using the simulated hydrogen ion density, hydrogen neutral density, electron density, electron temperature and neutral/ion temperature. Molecular components to the Balmer line emission are neglected, which are negligible for n>3 Balmer lines [52-56]. The Balmer line shape is modelled at each SOLPS grid cell by convolving the experimentally measured instrumental function with a Doppler broadening component and a Stark line shape ([6] and section 2), using the local SOLPS-simulated electron densities, electron temperatures and neutral/ion temperatures; leading to a Balmer line emission spectral profile (ph m$^{-3}$ s$^{-1}$ pix$^{-1}$) for each grid cell. The viewing cone corresponding to each synthetic spectroscopic line of sight is discretised as multiple lines of sight, which is further discretised into multiple points along each line. The emissivity spectra at each point is that of the corresponding SOLPS grid cell is calculated based on the cell characteristics ($n_e$, $T_e$, $n_o$, ion temperature and hydrogen ion density). The synthetic Balmer line spectra (ph m$^{-2}$ s$^{-1}$ pix$^{-1}$) for the entire chord is then obtained by integrating along each chord and summing the spectra obtained for each chord. This synthetic diagnostic implementation has been verified against the CHERAB code [65, 66].

In this analysis, the n=7 Balmer line is used for the Stark density and the n=5,6 Balmer line brightness and line ratio are used, similar to the technique used for #56567 highlighted in section 6 and [11]. An estimate for the path length is obtained analogous to the experiment from the simulation's target ion flux and flux surfaces (section 3.2). For simplicity, the SOLPS-Eirene grid cells corresponding to the inner strike point have been omitted to prevent pollution in the synthetic spectra originating from the inner strike point. As there are only five separate SOLPS runs corresponding to 5 upstream densities, the techniques highlighted in section 3.2 (based on the assumption of a continuous $T_e$ decrease in the divertor) to improve the $F_{rec}$ determination in limiting regimes ($F_{rec}$ ($F_{exc}$) < 0.1 – section 3.2) have not been used as they require a smooth evolution of the temperature in the divertor such that it can be assumed that the divertor temperatures are continuously decreasing (or at least not increasing). After employing the techniques in section 3 and 4, estimates of $I_L$, $P_{rad,L}^{H-exc}$, $CX_L$, $T_e^E$, $R_L$, $P_{rad,L}^{H-rec}$ for three of the five different simulations with 68% uncertainty margins are obtained along the outer divertor leg as shown in Figure 13c-p.

The above synthetic diagnostic 'measurements', based on generating and analysing spectra created from SOLPS local parameters, are compared to results obtained directly from local SOLPS values for the quantities of interest (e.g. radiation, ionization, recombination, …): line integrated quantities are obtained by employing the same line of sight geometry as the synthetic diagnostic and summing the contributions (mapped from the SOLPS grid to the diagnostic chords) over the diagnostic chordal path. Total synthetic diagnostic integrals of recombination/ionisation ($I_r$, $I_i$) are directly compared to the SOLPS output by summing the ionisation/recombination rates at every divertor grid cell that lies between the two outer spectroscopic lines of sight – Figure 13a-b. The density and $T_e^E$ are an exception in that it is directly obtained from SOLPS by computing an average $n_e$, $T_e^E$ along the chordal line of sight weighted by the local emissivity.

## 5.2 Results

The values of $I_i$ and $I_r$ obtained from the synthetic diagnostic are in good agreement with the direct result from SOLPS (<5% deviation). The large rise in uncertainty in the synthetic diagnostic analysis result for $I_i$ in cold divertor conditions is due to $F_{exc}$ being in a limiting regime ($F_{exc} < 0.1$). Good agreement between the synthetic diagnostic with direct SOLPS results is found for all line-integrated parameters ($F_{rec}$ (n=5) Figure 13c, d; $I_L$ Figure 13g, h; $R_L$ Figure 13i, j; Figure 13k, l) except the charge exchange to ionisation ratio ($CX_L / I_L$ Figure 13m, n). The deviation of the $CX_L/I_L$ ratio and the strong increase in the $I_L$ uncertainty near the target at the highest upstream densities occur when $F_{exc}$ is in its limiting regime ($F_{exc}$ (n=5) < 0.1). Such uncertainties, larger than the uncertainties during the experiment in the parameters estimated, from the excitation emission can be expected without applying the techniques in section 3.2. In addition, the two highest upstream density simulations investigated are more strongly detached than the experiment with higher recombination and lower $F_{exc}$.

We have also used the SOLPS model of TCV plasmas to examine the interpretation of local quantities (e.g. density; temperature) from chordal integrated emissivities through passive spectroscopy. To this end, the local ('slab') quantities inferred from DSS chordal measurements are compared to the emission-weighted averaged quantities along the chord through the SOLPS grid. A good agreement between the Stark density (from the synthetic diagnostic) and the emission-weighted $n_e$ (SOLPS) is shown (Figure 13e, f), indicating that the Stark density from the synthetic diagnostic (or, by implication, the direct analysis of DSS data) can be interpreted as the 'characteristic density' of the emission region. There is qualitative agreement (variation with core conditions) between the inferred $T_e^E$ from the synthetic diagnostic and the $T_e$ respective of the excitation emission region of the n=5 Balmer line.

The reason for poor quantitative agreement is the reduced sensitivity of the $T_e^E$ inference at larger $T_e$; the magnitude of the excitation emission becomes relatively more insensitive to the electron temperature. That is also evident from examination of the PDFs obtained and presented in Figure 11, showing a wide PDF for $T_e^E$ at higher temperatures. All of this indicates that the inferred local parameters can be considered 'characteristic' parameters of the emission region – or an emission-weighted-average value along the chord. Therefore, parameters obtained from the *total* Balmer line emission (such as the Stark density) can vary between different Balmer lines as their emission locations can vary. That location, however, for the excitation/recombination emission *separately* is the same. Thus, local quantities inferred from the excitation and recombination emission *separately* (such as $T_e^E$, $T_e^R$) are nearly identical across Balmer lines. This also implies that the different emission location of Balmer lines arises from their different sensitivities to recombinative emission; e.g. their different $F_{rec}$ (n). The separation of the excitation/recombination emission through $F_{rec}$ makes the analysis, therefore, less sensitive to line integration effects and opens improved ways of analysing the Balmer line series: multiple Balmer lines may not be linkable to a *single* $T_e$ due to line-integration effects; however, they may be linked to a *single* excitation/recombination temperature.

*Figure 13: Comparison between results from a synthetic diagnostic (right) on SOLPS-Eirene discharges and the direct SOLPS results (left). a, b) Volumetrically integrated ionisation and recombination rates as function of upstream density. c-p Profiles of local and line-integrated quantities along the outer divertor leg. Line-integrated profiles: c, d) $F_{rec}$ (n=5). g, h) ionisation rate. i,j) recombination rate. k, l) hydrogenic radiation (excitation). m, n)*

*charge exchange to ionisation ratio. Local quantities (note that the left-hand column are line-averaged weighted by the respective ($n_e^{Stark}$ – n=7; $T_e^E$ – n=5) Balmer line local emissivity): e, f) $n_e$ (direct result) and Stark inference (synthetic diagnostic). o,p) excitation emission-weighted temperature (direct result) and $T_e^E$ obtained from synthetic diagnostic analysis.*

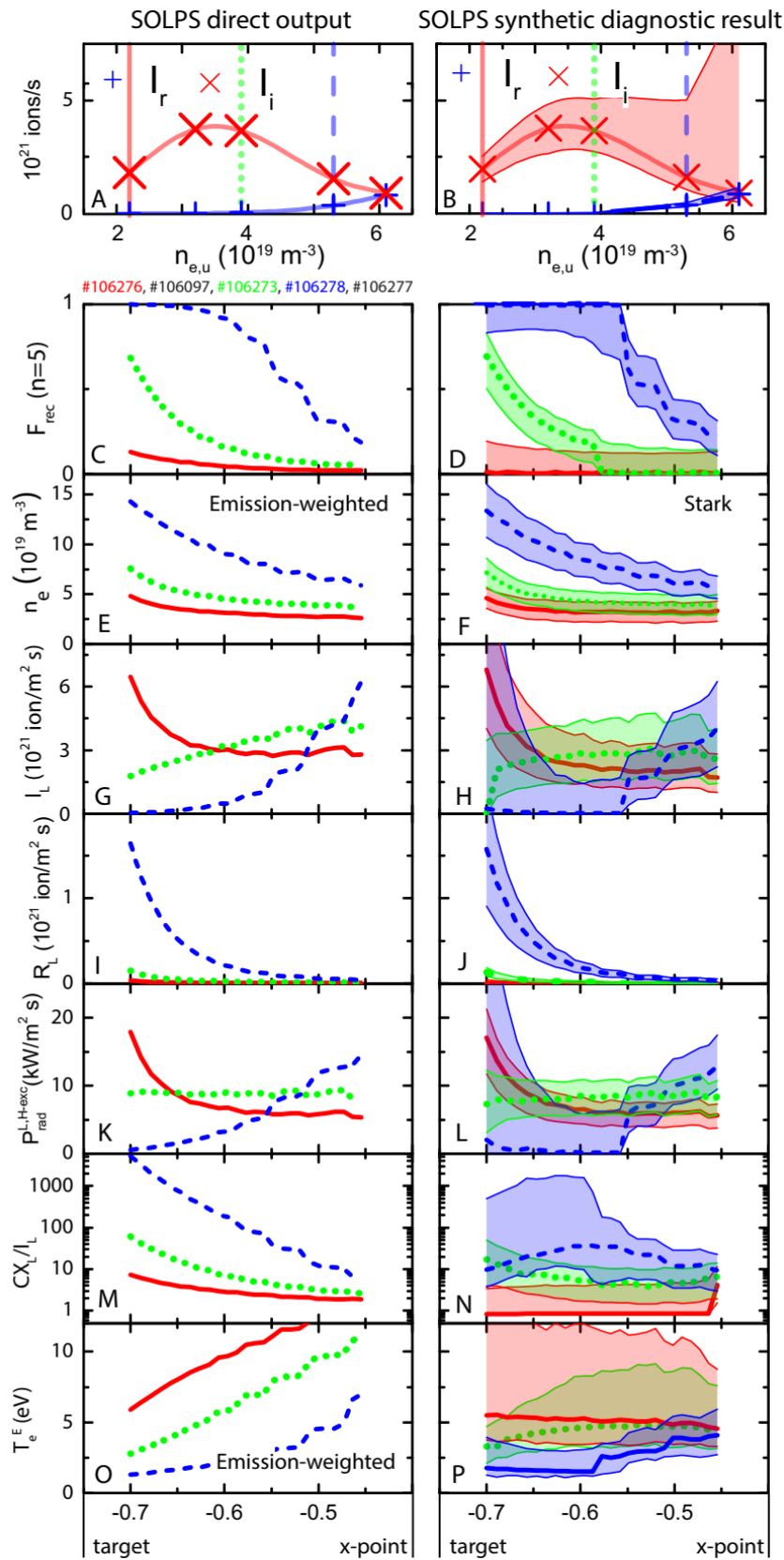

In summary, we find that although several assumptions are made in the analysis; the deviation of the analysis results from what their 'reality' (SOLPS in this case) is negligible compared to the uncertainty of the analysis. Hence, the analysis, particularly the line-integrated (such as $I_L$, $R_L$) and divertor-integrated (such as $I_i$ and $I_r$) ones, appears to be robust against line integration effects and assumptions regarding $Z_{eff}$. Simpler investigations of this analysis to ascertain the influence of line integration effects and $Z_{eff}$ on the analysis, irrespective of SOLPS simulations by using assumed a priori profiles and $Z_{eff}$ measurements, can be found by the author in [6, 10].

# 6. Experimental analysis illustration highlighting ion current loss during detachment through power limitation

An example of the application of the analysis techniques in this paper to obtain information on divertor particle and power balance during a characteristic Ohmic L-mode density ramp discharge (#56567) is shown in figure 14. That discharge is discussed in more detail in [10, 11], including detailed comparisons against the SOLPS simulations presented in section 5 as well as analytic models.

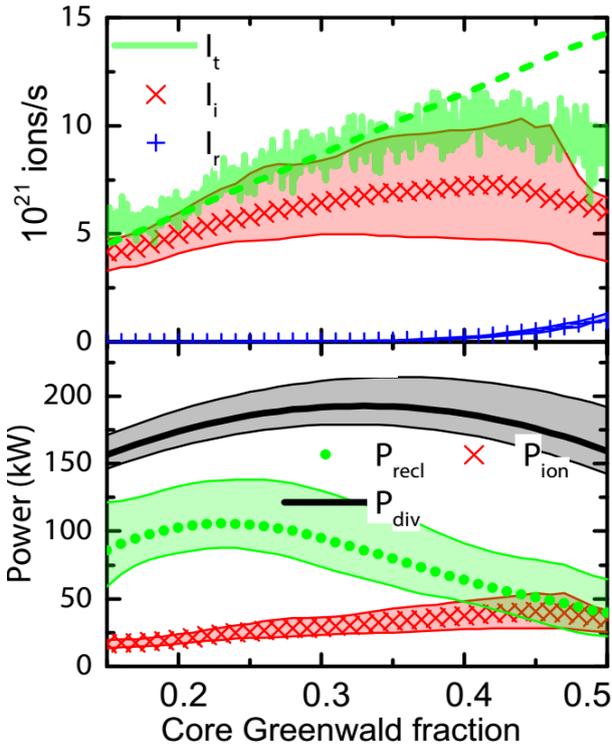

Figure 14: Power and particle balance of #56567 as function of core Greenwald fraction. a: Divertor particle balance comparing the ion target flux measured by Langmuir probes ($I_t$) against the divertor ionisation rate ($I_i$) and volumetric recombination rate ($I_r$) obtained through spectroscopic analysis. b: Divertor power balance comparing the power entering the divertor ($P_{div}$) against the power entering the recycling region ($P_{recl}$) (e.g. the power entering the divertor minus impurity radiation) and the power required for ionisation ($P_{ion}$) obtained spectroscopically.

We compare the total ion current (ion/s) reaching the target ($I_t$), measured by divertor Langmuir probes [67], to divertor ion sources (e.g. ionisation - $I_i$) and sinks (e.g. volumetric recombination - $I_r$) shown for a typical core plasma density ramp discharge in Fig. 14 a. The ion target current first increases linearly during the attached phase, after which $I_t$ begins to deviate from the linear trend (dashed line) at the detachment onset, ultimately rolling over [10, 11]. $I_t$ is quantitatively matched by the ion source: the flattening/roll-over of $I_t$ is most likely caused by a decrease in the ion source ($I_i$) rather than an increased ion sink from recombination ($I_r$), which is relatively small and only reaches relevant levels during deeper detachment when the target temperature attains values ≤1 eV.

Divertor power balance during a core density ramp, shown in figure 14b, indicates that hydrogenic ionisation related power losses $P_{ion}$ increase as the ion source increases. The power entering the divertor $P_{div}$ is significantly larger and remains roughly constant. Hydrogenic power losses, in this case, are significantly smaller than the total radiation – suggesting divertor impurity radiation dominates over hydrogenic radiation. Divertor impurity radiation ($P_{rad}^{imp}$) (which is estimated by subtracting hydrogenic radiation – estimated through spectroscopic analysis – from the total measured radiation by bolometry [10, 11]) continually lowers the power entering the ionisation region, $P_{recl} = P_{div} - P_{rad}^{imp}$ [9-11, 14] until it approaches the power loss associated with ionisation during detachment.

# 7. Discussion

## 7.1. Applicability to other fusion devices

The Balmer line analysis techniques developed in this paper have been applied to the TCV tokamak. However, TCV employs an open divertor with relatively low power crossing the separatrix (leading to relatively long scale lengths in the divertor) and operates at relatively low electron density. The question is whether the analysis techniques, described herein, will be more widely applicable to other devices. There are three main concerns for this: 1) other devices may have more complicated viewing geometry leading to more complicated emission profiles along the line of sights; 2) the neutral fraction – at the excitation emission region – may not be 'relatively' constant along each line of sight as assumed here, which would not lead to a clear transition from excitation to recombinative emission as the plasma cools down (section 3.6.3, Fig. 9); and 3) other devices may operate at higher densities/lower temperatures; which would drive up the amount of recombinative emission, complicating the extraction of excitation emission in strongly detached regimes.

Preliminary investigations using detached MAST-U SOLPS [68] simulations (like section 5) with/without $N_2$ seeding have shown that the analysis in this paper can be validly applied. Although ionisation estimates could not be obtained near the target in the strongest detached states (where target temperatures as low as 0.2 eV were obtained), most of the ionisation was still correctly detected through the use of synthetic spectroscopic analysis of the various planned chordal views. The viewing geometry for a MAST-U Super-X divertor and the chordal integral is, however, significantly more complicated than on TCV. For instance, the emission profile along the line of sight can be hollow and can include an excitation region surrounded by two separate recombination regions. This was not found to be an issue through the synthetic diagnostic analysis and provides confidence that the analysis described herein are sufficiently robust against line-integration effects, resolving the first point of concern for future application.

As explained in section 3.6.3, one assumes that the neutral fraction's dependence on the electron temperature is such that it does not change the relation between $F_{rec}$ and the Balmer line ratio. However, if instead one uses a neutral fraction that *is* dependent on temperature and the ion residence time, $\tau$, $n_o/n_e (T_e,\tau)$, the relationship between $F_{rec}$ and the Balmer line ratio does not change significantly; only minor changes are obtained when the neutral fraction rises above 1, which is larger than expectations based on JET results [43] – see section 3.6.3. Therefore, the second point of concern mentioned above does not seem to be an issue for the applicability of these techniques to future devices.

The third point of concern implies that for high density devices, such as ITER, DEMO and C-Mod, the Balmer line emission during detached operation (assuming <1.5 eV temperatures), could be too strongly dominated by recombination to uncover the excitation component. This would be a complication unless the recombination/ionisation regions are clearly separated between different lines of sight – which also has repercussions for the viewing geometry requirements. Considering that the scale lengths for such high power, high density devices are expected to be shorter than TCV (particularly normalized to the size of the device), it is likely that the ionisation/recombination regions are more localised than on TCV. If (significant) excitation/recombination regions lie on a single line of sight, the Balmer line – for such high density/low temperature regimes – would only give information on recombination and to obtain excitation/ionisation information, Lyman series measurements may have to be employed [31]. On the other hand, impurity seeded discharges, such as $N_2$ seeded discharges, generally do not reach as low a temperature during detachment compared to density ramp experiments [10, 68] as in the example shown in this paper. Another concern is the fact that opacity is ignored in the above analysis [17, 31]: opacity increases with $n_0\delta$ where $n_0$ is the neutral density and $\delta$ is the pathlength through the high $n_0$ region (can increase with machine/divertor size). Opacity could lead to a modification in the observed Balmer line intensity [17, 31] and could affect the ionisation balance [31, 69-74]. Thus, larger and higher density machines could introduce a limit to the maximum $n_0\delta$ at which this analysis could be applied.

## 7.2. The relation of 'recombinative dominated emission' to the actual recombination rate

One important point of our analysis techniques and accompanying discussion in section 3.4 is that Balmer line ratios do *not* provide information on the actual dominance of recombination *reactions* over ionisation; but instead information on the dominance of recombination *emission* over excitation emission.

This discussion has implications for the common use of line ratios to indicate the 'dominance' of recombination [7, 15, 34, 38, 75, 76]. For such an analysis, the line ratios used to quantify this recombination "dominance" correspond to $F_{rec} \sim 1$ in Figure 3. However, as shown in Figure 4, the "dominance" of higher-n Balmer line emission ($F_{rec}$ (n=6,7) > 0.9) commences at recombination to ionisation rate ratios of 1-10% (depending on the density – Figure 4). Therefore, even if higher-n Balmer line emission is dominated by recombination, the ionisation rate can still be much higher than the recombination rate! Although line ratios can be employed to gauge whether recombination is present and whether the Balmer line emission of a particular transition is dominated by recombination, they, by themselves, do not provide direct information on the magnitude of volumetric recombination and on the value of the recombination to ionisation ratio. Instead, quantitative calculations must be performed to infer both the magnitude of the ionisation and recombination rates.

Additionally, Balmer line ratio trends can provide information on the behaviour of the neutrals in the divertor. The discussion in section 3.6.3 has indicated that the *trend* of Balmer line ratios during conditions where the divertor temperature is continuously decreased (e.g. seeding scans and/or density ramps) provides information on the neutral fraction in the divertor, which in turn can be used as a diagnostic to investigate how far the neutral dynamics deviate from a local ionisation/recombination equilibrium.

## 8. Conclusion

A novel approach of analysing the Balmer line series has enabled the simultaneous inference of ionisation/recombination rates as well as hydrogenic power losses associated with ionisation: giving rise to a full power/particle balance investigation of the divertor. Techniques in this work have been developed to use the Balmer line ratio to quantitatively separate excitation/recombination emission. Analysing each contribution individually then leads to ionisation/recombination rate estimates using a robust technique which is relatively insensitive to line integration effects – verified using a synthetic diagnostic approach on SOLPS simulations.

## Acknowledgments

This work has been carried out within the framework of the EUROfusion Consortium and has received funding from the Euratom research and training programme 2019–2020 under Grant Agreement No. 633053. The views and opinions expressed herein do not necessarily reflect those of the European Commission. This work was supported in part by the Swiss National Science Foundation. The PhD research of K. Verhaegh was supported by funding from the University of York and the Swiss National Science Foundation. B. Lipschultz was funded in part by the Wolfson Foundation and UK Royal Society through a Royal Society Wolfson Research Merit Award as well as by the RCUK Energy Programme (EPSRC grant number EP/I501045).

## References

1. Loarte, A., et al., *Chapter 4: Power and particle control.* Nuclear Fusion, 2007. **47**(6): p. S203-S263.
2. Stangeby, P.C., *Basic physical processes and reduced models for plasma detachment.* Plasma Physics and Controlled Fusion, 2018. **60**(4): p. 044022.
3. Reimerdes, H., et al., *TCV experiments towards the development of a plasma exhaust solution.* Nuclear Fusion, 2017. **57**(12): p. 126007.
4. Theiler, C., et al., *Results from recent detachment experiments in alternative divertor configurations on TCV.* Nuclear Fusion, 2017. **57**(7): p. 072008.
5. Krasheninnikov, S.I. and A.S. Kukushkin, *Physics of ultimate detachment of a tokamak divertor plasma.* Journal of Plasma Physics, 2017. **83**(5): p. 155830501.
6. Verhaegh, K., et al., *Spectroscopic investigations of divertor detachment in TCV.* Nuclear Materials and Energy, 2017. **12**: p. 1112-1117.
7. Harrison, J.R., et al., *Detachment evolution on the TCV tokamak.* Nuclear Materials and Energy, 2017. **12**: p. 1071-1076.
8. Lipschultz, B., B. LaBombard, J.L. Terry, C. Boswell, and I.H. Hutchinson, *Divertor physics research on Alcator C-Mod.* Fusion Science and Technology, 2007. **51**(3): p. 369-389.
9. Lipschultz, B., et al., *The role of particle sinks and sources in Alcator C-Mod detached divertor discharges.* Physics of Plasmas, 1999. **6**(5): p. 1907-1916.
10. Verhaegh, K., *Spectroscopic Investigations of detachment on TCV.* 2018, PhD Thesis, University of York Available from: http://etheses.whiterose.ac.uk/22523/.
11. Verhaegh, K., et al., *An improved understanding of the roles of atomic processes and power balance in divertor target ion current loss during detachment.* Nuclear Fusion, 2019
12. Dudson, B., J. Allen, T. Body, B. Chapman, C. Lau, L. Townley, D. Moulton, J. Harrison, and B. Lipschultz, *The role of particle, energy and momentum losses in 1D simulations of divertor detachment.* Plasma Phys. Control. Fusion, 2019. **61**(065008)
13. Pshenov, A.A., A.S. Kukushkin, and S.I. Krasheninnikov, *Energy balance in plasma detachment.* Nuclear Materials and Energy, 2017. **12**: p. 948-952.
14. Krasheninnikov, S.I., A.S. Kukushkin, and A.A. Pshenov, *Divertor plasma detachment.* Physics of Plasmas, 2016. **23**(5): p. 055602.
15. McCracken, G.M., M.F. Stamp, R.D. Monk, A.G. Meigs, J. Lingertat, R. Prentice, A. Starling, R.J. Smith, and A. Tabasso, *Evidence for volume recombination in jet detached divertor plasmas.* Nuclear Fusion, 1998. **38**(4): p. 619-629.
16. Lumma, D., J.L. Terry, and B. Lipschultz, *Radiative and three-body recombination in the Alcator C-Mod divertor.* Physics of Plasmas, 1997. **4**(7): p. 2555-2566.
17. Terry, J.L., B. Lipschultz, A.Y. Pigarov, S.I. Krasheninnikov, B. LaBombard, D. Lumma, H. Ohkawa, D. Pappas, and M. Umansky, *Volume recombination and opacity in Alcator C-Mod divertor plasmas.* Physics of Plasmas, 1998. **5**(5): p. 1759-1766.
18. Wenzel, U., K. Behringer, A. Carlson, J. Gafert, B. Napiontek, and A. Thoma, *Volume recombination in divertor I of ASDEX Upgrade.* Nuclear Fusion, 1999. **39**(7): p. 873.

19. Lipschultz, B., J.L. Terry, C. Boswell, A. Hubbard, B. LaBombard, and D.A. Pappas, *Ultrahigh densities and volume recombination inside the separatrix of the Alcator C-Mod tokamak.* Physical Review Letters, 1998. **81**(5): p. 1007-1010.
20. Terry, J.L., B. Lipschultz, X. Bonnin, C. Boswell, S.I. Krasheninnikov, A.Y. Pigarov, B. LaBombard, D.A. Pappas, and H.A. Scott, *The experimental determination of the volume recombination rate in tokamak divertors.* Journal of Nuclear Materials, 1999. **266-269**: p. 30-36.
21. Post, D.E., *A Review of Recent Developments in Atomic Processes for Divertors and Edge Plasmas.* Journal of Nuclear Materials, 1995. **220**: p. 143-157.
22. Wising, F., S. Krasheninnikov, D. Sigmar, D. Knoll, T. Rognlien, B. LaBombard, B. Lipschultz, and G. McCracken, *Simulation of plasma flux detachment in Alcator C-Mod and ITER.* Journal of nuclear materials, 1997. **241**: p. 273-277.
23. Loarte, A., *Understanding the edge physics of divertor experiments by comparison of 2D edge code calculations and experimental measurements.* Journal of Nuclear Materials, 1997. **241-243**: p. 118-134.
24. Krasheninnikov, S., A.Y. Pigarov, D. Knoll, B. LaBombard, B. Lipschultz, D. Sigmar, T. Soboleva, J. Terry, and F. Wising, *Plasma recombination and molecular effects in tokamak divertors and divertor simulators.* Physics of Plasmas, 1997. **4**(5): p. 1638-1646.
25. Borrass, K., R. Schneider, and R. Farengo, *A scrape-off layer based model for Hugill-Greenwald type density limits.* Nuclear fusion, 1997. **37**(4): p. 523.
26. Pitts, R.A., et al., *Divertor geometry effects on detachment in TCV.* Journal of Nuclear Materials, 2001. **290**: p. 940-946.
27. Fil, A.M.D., B.D. Dudson, B. Lipschultz, D. Moulton, K.H.A. Verhaegh, O. Fevrier, and M. Wensing, *Identification of the primary processes that lead to the drop in divertor target ion current at detachment in TCV.* Contributions to plasma physics, 2017. **58**(6-8)
28. Reimold, F., M. Wischmeier, S. Potzel, L. Guimarais, D. Reiter, M. Bernert, M. Dunne, T. Lunt, A.U. Team, and E. Mst1Team, *The high field side high density region in SOLPS-modeling of nitrogen-seeded H-modes in ASDEX Upgrade.* Nuclear Materials and Energy, 2017. **12**: p. 193-199.
29. Wischmeier, M., *Simulating divertor detachment in the TCV and JET tokamaks.* 2005,PhD Thesis, EPFL DOI: 10.5075/epfl-thesis-3176.
30. Monk, R.D., et al., *Interpretation of ion flux and electron temperature profiles at the JET divertor target during high recycling and detached discharges.* Journal of Nuclear Materials, 1997. **241-243**: p. 396-401.
31. Lomanowski, B., et al., *Spectroscopic investigation of N2 and Ne seeded induced detachment in JET ITER-like wall.* Nuclear Materials and Energy, 2019,. **20**(100676)
32. Lomanowski, B., et al. *Integrated spectroscopic analysis of L-mode detachment in JET-ILW*. in *44th EPS Conference on Plasma Physics*. 2017. Belfast.
33. Verhaegh, K., et al. *Spectroscopic investigation of ion sources/sinks during TCV detachment*. in *44th EPS Conference on Plasma Physics*. 2017. Belfast DOI: 10.13140/RG.2.2.29588.50568/3.
34. Potzel, S., M. Wischmeier, M. Bernert, R. Dux, H.W. Muller, A. Scarabosio, and A.U. Team, *A new experimental classification of divertor detachment in ASDEX Upgrade.* Nuclear Fusion, 2014. **54**(1): p. 013001.
35. Lipschultz, B., J.L. Terry, C. Boswell, S.I. Krasheninnikov, B. LaBombard, and D.A. Pappas, *Recombination and ion loss in C-Mod detached divertor discharges.* Journal of Nuclear Materials, 1999. **266-269**: p. 370-375.
36. Henderson, S.S., et al., *Determination of volumetric plasma parameters from spectroscopic N II and N III line ratio measurements in the ASDEX Upgrade divertor.* Nuclear Fusion, 2018. **58**(1): p. 016047.
37. Potzel, S., *Experimental classification of divertor detachment.* 2012,PhD Thesis.

38. Stangeby, P., *The plasma boundary of magnetic fusion devices.* The Plasma Boundary of Magnetic Fusion Devices. Series: Series in Plasma Physics, ISBN: 978-0-7503-0559-4. Taylor & Francis, Edited by Peter Stangeby, vol. 7, 2000. **7**
39. Verhaegh, K., *kevinverhaegh/ionrec: Ionisation/recombination spectroscopy analysis code for tokamak plasmas; Zenodo: 10.5281/zenodo.2573219.* 2019 DOI: 10.5281/zenodo.2573219.
40. *OPEN-ADAS; Open - Atomic Data Analysis Structure.* Available from: http://http://open.adas.ac.uk/.
41. Summers, H.P., W.J. Dickson, M.G. O'Mullane, N.R. Badnell, A.D. Whiteford, D.H. Brooks, J. Lang, S.D. Loch, and D.C. Griffin, *Ionization state, excited populations and emission of impurities in dynamic finite density plasmas: I. The generalized collisional–radiative model for light elements.* Plasma Physics and Controlled Fusion, 2006. **48**(2): p. 263-293.
42. Mullane, M.O. *Generalised Collisional Radiative data for hydrogen: ADAS Manual*. 2013; Available from: http://www.adas.ac.uk/notes/adas_c13-01.pdf.
43. Lomanowski, B.A., A.G. Meigs, R.M. Sharples, M. Stamp, C. Guillemaut, and J. Contributors, *Inferring divertor plasma properties from hydrogen Balmer and Paschen series spectroscopy in JET-ILW.* Nuclear Fusion, 2015. **55**(12): p. 123028.
44. Stehlé, C. and R. Hutcheon, *Extensive tabulations of Stark broadened hydrogen line profiles.* Astronomy and Astrophysics Supplement Series, 1999. **140**(1): p. 93-97.
45. Rosato, J., Y. Marandet, and R. Stamm, *A new table of Balmer line shapes for the diagnostic of magnetic fusion plasmas.* Journal of Quantitative Spectroscopy and Radiative Transfer, 2017. **187**: p. 333-337.
46. Rosato, J., N. Kieu, M. Meireni, R. Sheeba, M. Koubiti, Y. Marandet, R. Stamm, K. Verhaegh, and B. Duval, *Stark broadening of Balmer lines with low and moderate quantum number in dense divertor plasmas.* Contrib. Plasma Phys., 2017
47. Reimold, F., M. Wischmeier, M. Bernert, S. Potzel, A. Kallenbach, H.W. Muller, B. Sieglin, U. Stroth, and A.U. Team, *Divertor studies in nitrogen induced completely detached H-modes in full tungsten ASDEX Upgrade.* Nuclear Fusion, 2015. **55**(3): p. 033004.
48. Reiter, D. *http://www.eirene.de*. 2019.
49. Reiter, D., M. Baelmans, and P. Börner, *The EIRENE and B2-EIRENE Codes.* Fusion Science and Technology, 2005. **47**(2): p. 172-186.
50. Wesson, J., *Tokamaks*. Vol. 149. 2011: Oxford University Press.
51. Sakamoto, M., et al., *Molecular activated recombination in divertor simulation plasma on GAMMA 10/PDX.* Nuclear Materials and Energy, 2017. **12**: p. 1004-1009.
52. Groth, M., et al., *EDGE2D-EIRENE predictions of molecular emission in DIII-D high-recycling divertor plasmas.* Nuclear Materials and Energy, 2019. **19**: p. 211-217.
53. Wünderlich, D. and U. Fantz, *Evaluation of State-Resolved Reaction Probabilities and Their Application in Population Models for He, H, and H2.* Atoms, 2016. **4**(4)
54. Fantz, U., *Emission spectroscopy of hydrogen molecules in technical and divertor plasmas.* Contributions to Plasma Physics, 2002. **42**(6-7): p. 675-684.
55. Fantz, U., D. Reiter, B. Heger, and D. Coster, *Hydrogen molecules in the divertor of ASDEX Upgrade.* Journal of Nuclear Materials, 2001. **290**: p. 367-373.
56. Hollmann, E.M., S. Brezinsek, N.H. Brooks, M. Groth, A.G. McLean, A.Y. Pigarov, and D.L. Rudakov, *Spectroscopic measurement of atomic and molecular deuterium fluxes in the DIII-D plasma edge.* Plasma Physics and Controlled Fusion, 2006. **48**(8): p. 1165.
57. Maurizio, R., S. Elmore, N. Fedorczak, A. Gallo, H. Reimerdes, B. Labit, C. Theiler, C. Tsui, W. Vijvers, and T. Team, *Divertor power load studies for attached L-mode single-null plasmas in TCV.* Nuclear Fusion, 2017. **58**(1): p. 016052.
58. Perek, A., et al., *MANTIS: a real-time quantitative multispectral imaging system for fusion plasmas.* Rev Sci Instrum, 2019, submitted
59. Wischmeier, M., et al., *The influence of molecular dynamics on divertor detachment in TCV.* Contributions to Plasma Physics, 2004. **44**(1-3): p. 268-273.

60. Carolan, P.G. and V.A. Piotrowicz, *The behaviour of impurities out of coronal equilibrium.* Plasma Physics, 1983. **25**(10): p. 1065-1086.
61. Kallenbach, A., et al., *Impurity seeding for tokamak power exhaust.* Plasma Physics and Controlled Fusion, 2013. **55**(12): p. 124041.
62. Botev, Z.I., J.F. Grotowski, and D.P. Kroese, *Kernel density estimation via diffusion.* The Annals of Statistics, 2010. **38**(5): p. 2916-2957.
63. Bowman, C., *Applications of Bayesian Probability Theory in Fusion Data Analysis.* 2016, PhD Thesis, University of York.
64. Cowles, M.K., *Applied Bayesian statistics: with R and OpenBUGS examples*. Vol. 98. 2013: Springer Science & Business Media.
65. Carr, M., A. Meakins, M. Bernert, P. David, C. Giroud, J. Harrison, S. Henderson, B. Lipschultz, F. Reimold, and E.M. Team, *Description of complex viewing geometries of fusion tomography diagnostics by ray-tracing.* Review of Scientific Instruments, 2018. **89**(8): p. 083506.
66. Giroud, C., A. Meakins, M. Carr, A. Baciero, and C. Bertrand, *CHERAB spectroscopy modelling framework*. 2018, Zenodo.
67. Février, O., C. Theiler, H.D. Oliveira, B. Labit, N. Fedorczak, and A. Baillod, *Analysis of wall-embedded Langmuir probe signals in different conditions on the Tokamak à Configuration Variable.* Rev Sci Instrum, 2018. **89**(5): p. 053502.
68. Myatra, O., D. Moulton, A. Fil, B. Dudson, and B. Lipschultz. *Taming the flame: Detachment access and control in MAST-U Super-X*. in *Plasma Surface Interactions*. 2018. Princeton.
69. Marenkov, E., S. Krasheninnikov, and A. Pshenov, *Multi-level model of radiation transport in inhomogeneous plasma.* Contributions to Plasma Physics, 2018. **58**(6-8): p. 570-577.
70. Rosato, J., Y. Marandet, D. Reiter, and R. Stamm, *Development of a hybrid kinetic-fluid model for line radiation transport in magnetic fusion plasmas.* High Energy Density Physics, 2017. **22**: p. 73-76.
71. Hoshino, K., K. Sawada, R. Idei, S. Tokunaga, N. Asakura, K. Shimizu, and N. Ohno, *Photon Trapping Effects in DEMO Divertor Plasma.* Contributions to Plasma Physics, 2016. **56**(6-8): p. 657-662.
72. Kotov, V., D. Reiter, A.S. Kukushkin, H.D. Pacher, P. Börner, and S. Wiesen, *Radiation Absorption Effects in B2-EIRENE Divertor Modelling.* Contributions to Plasma Physics, 2006. **46**(7-9): p. 635-642.
73. Pshenov, A., A.S. Kukushkin, E. Marenkov, and S.I. Krasheninnikov, *On the role of hydrogen radiation absorption in divertor plasma detachment.* Nuclear Fusion, 2019
74. Scott, H.A. and M.L. Adams, *Incorporating Line Radiation Effects into Edge Plasma Codes.* Contributions to Plasma Physics, 2004. **44**(1-3): p. 51-56.
75. Ramasubramanian, N., R. König, Y. Feng, L. Giannone, P. Grigull, T. Klinger, K. McCormick, H. Thomsen, U. Wenzel, and W.A.S.T. the, *Characterization of the island divertor plasma of W7-AS stellarator in the deeply detached state with volume recombination.* Nuclear Fusion, 2004. **44**(9): p. 992.
76. Terry, J.L. and M.L. Reinke, *Diagnostic tools for studying divertor detachment: bolometry, spectroscopy, and thermography for surface heat-flux.* Plasma Physics and Controlled Fusion, 2017. **59**(4): p. 044004.